\documentclass[sigconf]{acmart}

\usepackage{import}

\usepackage{graphicx} 
\graphicspath{{./Figures/}} 
\usepackage{subfig}

\usepackage{siunitx}

\usepackage{tabularx}
\usepackage{multirow}
\usepackage{array}

\usepackage{enumitem}
\usepackage{xcolor} 
\usepackage{soul}
\usepackage{caption}
\usepackage{xtab}
\usepackage{url} 

\usepackage{algorithmic}

\usepackage[]{verbatim}

\AtBeginDocument{%
  \providecommand\BibTeX{{%
    \normalfont B\kern-0.5em{\scshape i\kern-0.25em b}\kern-0.8em\TeX}}}

\setcopyright{none}
\copyrightyear{2020}
\acmDOI{}
\acmISBN{}
\acmPrice{}
\acmConference[eprint'2020]{arXiv}{}{US}
\acmYear{2020}

\begin{document}

\title{RanStop: A Hardware-assisted Runtime Crypto-Ransomware Detection Technique}

\author{Nitin Pundir}
\email{nitin.pundir@ufl.edu}
\affiliation{%
  \institution{University of Florida, US}
}

\author{Mark Tehranipoor}
\email{tehranipoor@ece.ufl.edu}
\affiliation{%
  \institution{University of Florida, US}
}

\author{Fahim Rahman}
\email{fahimrahman@ece.ufl.edu}
\affiliation{%
  \institution{University of Florida, US}
}


\begin{abstract}
Among many prevailing malware, crypto-ransomware poses a significant threat as it financially extorts affected users by creating denial of access via unauthorized encryption of their documents as well as holding their documents hostage and financially extorting them. This results in millions of dollars of annual losses of the infected victims worldwide. Multiple variants of ransomware are growing in number with capabilities of evasion from many anti-viruses and software-only malware detection schemes that rely on static execution signatures. In this paper, we propose a hardware-assisted scheme, called RanStop, for early detection of crypto-ransomware infection in commodity processors. Specifically, RanStop leverages the information of hardware performance counters (HPCs) embedded in the performance monitoring unit (PMU) in modern processors to observe micro-architectural event sets and detects known and unknown crypto-ransomware variants. In this paper, we train a recurrent neural network-based machine learning architecture (RNN) using long short-term memory (LSTM) model for analyzing micro-architectural events in the hardware domain when executing multiple variants of ransomware as well as benign programs (goodware). We create timeseries to develop intrinsic statistical features using the information of related HPCs and improve the detection accuracy of RanStop and reduce noise by via LSTM and global average pooling (GAP). As an early detection scheme, RanStop can accurately and quickly identify ransomware within 2\si{\milli\second} from the start of the program execution by analyzing HPC information collected for 20 timestamps each 100\si{\micro\second} apart. This detection time is too early for a ransomware to make any significant damage, if none. Moreover, validation against benign programs with behavioral (sub-routine-centric) similarity with that of a crypto-ransomware shows that RanStop can detect ransomware with an average of \si{97\percent} accuracy for fifty random trials. 
\end{abstract}


\maketitle

\section{Introduction}\label{sec:introduction}


Security vulnerabilities in modern computing systems for stand-alone and networked applications have given rise to numerous cyber attacks and malware that can cause privacy breach and data loss, compromise critical infrastructures and national security, cause financial damage, and more. McAfee quarterly threat report (Jan-Mar 2017) reveals that 176 new cyber threats are being emerging every minute \cite{mcafee2017threat}. Among such cyber threats and attacks, \textit{ransomware} has gained much attention due to its malicious nature and subsequent exploitation, loss of data, and financial loss \cite{hutcherson2018ransomware,vamosi2018wannacry,o2012ransomware,liptak_2017}.

Ransomware, comprising of the words `ransom' and `malware', is a class of malware which asks for ransom/money from the victim via anonymous payment mechanisms by holding the system/files as hostage, in exchange for restoring the hijacked functionality. Since the majority of the ransomware use cryptographic encryption-decryption processes and key (password) exchange protocols for `locking' (and later releasing) the data stored in the infected system, this class is also known as crypto-ransomware (as oppose to the locker-ransomware that causes a denial-of-service on the system to restrict operational access by the user) \cite{scaife2016cryptolock, sgandurra2016automated}. Although existed for more than a decade, a massive rise of crypto-ransomware is observed in the past few years as more devices with inadequate protection are being connected to the global network and its multiple variants with stealthy nature are being fast spread out too often. So far, the family of crypto-ransomware has caused substantial global financial loss and is continuing to cost tens of millions of dollars in consumer losses annually \cite{vamosi2018wannacry,scaife2016cryptolock}. The actual cost incurred by the ransomware is believed to be much higher since numerous incidents remain unreported from the victims. In addition to financial loss, crypto-ransomware attacks, especially on business organizations, law enforcement agencies, and even entertainment groups, have resulted into loss of critical data, private information, valuable documents, work hours, and services \cite{arnold2014tennessee,hutcherson2018ransomware,liptak_2017,radiohead2019ok}. Ransomware is considered as the main reason of 39\% of malware-related data breaches in 2018, according to Verizon's annual report \cite{alison2018ransomware}. As such, crypto-ransomware represents a major threat to all classes of users in the modern world.

In the past few years, researchers have put much emphasis on detecting and preventing ransomware. The majority of the proposed techniques are based on static software-centric set-up which typically utilizes static methods such as signature (template) matching in the program control flow, searching for dominant features of malicious ransomware-like activities, and monitoring high-level (software) execution, API/system calls, or data to detect potentially anomalous behavior \cite{sgandurra2016automated,scaife2016cryptolock,kharraz2016unveil,andronio2015heldroid}. Unfortunately, such static and software-only schemes, including commercial anti-viruses, often fail to provide a comprehensive security against ever-growing ransomware. Two prime examples of ransomware evading existing software-only mechanisms are -- (1) the latest ransomware attacks on the city of Atlanta's online systems shutting down the activities for more than six days \cite{hutcherson2018ransomware}; and (2) the WannaCry malware infecting many business organizations in over 150 different countries\cite{vamosi2018wannacry,liptak_2017}. In both cases, systems under attack presumably had adequate protection against potential crypto-ransomware and denial of service attacks. Software-only schemes suffer from the following intrinsic limitations:

\begin{itemize}
    \item 
    There are numerous different ransomware with distinct and obscure control flow signatures that can evade static signature-matching anti-malware schemes.
    
    \item 
    Software-based techniques often require binary signature for each variant of the ransomware (or, in general, malware). It imposes huge overhead on the database, i.e., the size of anti-malware updates, with ever-growing polymorphic and metamorphic variants of a class of malware \cite{bishop2005introduction}.
    
    \item 
    Even in the case of a successful detection, several existing schemes are too late (in time) at detecting a ransomware such that the victim system and important files (or directories) may already be maliciously corrupted (encrypted) and locked, where the possibility of the recovery is extremely rare.
    
    \item
    The static signature mapping can produce a high rate of false decisions (false positive/negative) which pose critical impact on the smooth operation of the system.
    
\end{itemize}

Therefore, it is apparent that the existing software-only techniques are not adequate to thwart crypto-ransomware attacks. To address the existing challenges and limitations, we propose a hardware-assisted runtime crypto-ransomware detection scheme, called \textit{RanStop}, for commodity computing platforms. RanStop leverages existing hardware performance counters (HPCs) in the performance monitoring units (PMUs), commonly available in recent generation processors \cite{intel64manual,armA9manual}, for runtime event-monitoring at the micro-architectural level and utilizes state-of-the-art machine learning technique to develop an advanced predictive model for an early and accurate detection of known and unknown variants of the crypto-ransomware family.

The main motivation behind developing our hardware-assisted crypto-ransomware detection technique, RanStop, comes from the fact that HPCs can collect multi-dimensional hardware event-traces and micro-architectural information during program execution with zero hardware modification. Although originally designed for performance monitoring, HPCs (and PMUs) can be intelligently used for security by analyzing whether a runtime event profile is malicious in nature. Monitoring the information of hardware micro-architectural events and developing ML-based machine learning predictive models provide us advantages over high-level software features as follows.


\begin{enumerate}
	
	\item 
	Being an integrated part of the hardware, HPCs operate transparently to any software running on the processor and collect targeted micro-architectural data irrespective to the program execution mode. Therefore, software-obfuscated or stealthy ransomware cannot evade them.
	 
	\item 
	It offers multi-dimensional information from a large set of micro-architectural sources. Therefore, acquired information can be utilized for multi-modal analysis techniques such as statistical analysis and machine learning techniques. 
	
	\item 
	Being an integrated part of the hardware, it is able to collect information significantly faster (often in \si{\micro\second} ranges) than any software-centric trace acquisition approach.
	
	\item 
	Developed ML model can be retrained with additional dataset for emerging ransomwares with little to no modification in the ML framework. Also, the developed ML model size is significantly smaller compared to the static signature-based database \cite{alam2019ratafia}.

\end{enumerate}

\begin{figure} [t] 
	\begin{center}
		\includegraphics[keepaspectratio,width=01\columnwidth, height=7.8in]{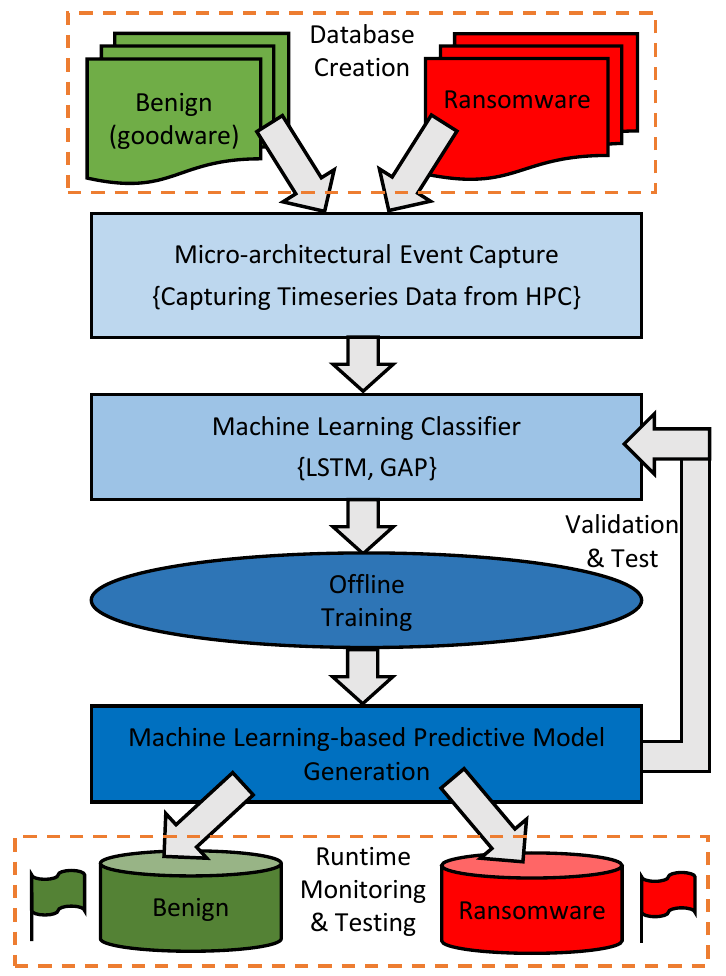}
		\caption[]{Primary workflow for RanStop.} \label{fig:BlockDiagram}
	\end{center}
\end{figure}
\vspace{0.05in}

The nature of cryptro-ransomware family itself shows certain dependency on various sub-routines (e.g., encryption and data movement) that can be common in regular benign programs (e.g., disk encryption goodwares) and cannot be accurately distinguished only via static signature analysis or control/data flow graph (CDFG) \cite{moussaileb2018ransomware, alam2019ratafia}. Our proposed RanStop technique can identify intrinsic dissimilarities between crypto-ransomware and goodware from the hardware activity signatures. RanStop, as shown in Figure \ref{fig:BlockDiagram}, first collects runtime micro-architectural event signatures for all HPC groups for both the ransomware and goodware. Using the collected hardware information, a machine learning model is then developed which takes a small runtime data and can detect ransomware with high accuracy. The contributions of our work are as follows.

\begin{enumerate}
	
	\item 
	RanStop offers an accurate and noise-free collection of hardware-domain micro-architectural activities, e.g., branch prediction success/miss, \si{L2} cache access, etc. for ongoing program (command) executions for detecting malicious event traces. This dynamic approach is applicable for both known and unknown Ransomware with minimal runtime and zero hardware overhead.
	 
	\item 
	RanStop utilizes the state-of-the-art machine learning (ML) techniques with intelligent feature selection scheme to accurately detect ransomware. We carry out extensive analysis for \si{80} crypto-ransomware using recurrent neural network (RNN) architecture using long short-term memory (LSTM) and global average pooling methods \cite{ hochreiter1997long,gers2002learning}. Our technique provides {\si{97\%}} prediction accuracy on average for selected hardware performance groups. 
	
	\item 
	RanStop offers significantly early-detection for ransomware by analyzing the micro-architectural data collected for 20 timestamps each \si{100\micro\second} apart from the start of the execution (2\si{\milli\second} in total). This allows to stop the malicious execution at a very early stage and protects the system and files long before undergoing significant, if not none, damage and data loss.

\end{enumerate}

The rest of the paper is organized as follows: Section \ref{sec:preliminaries} highlights prior work and relevant concepts. Section \ref{sec:methodology} describes the workflow of our proposed RanStop technique in detail. Section \ref{sec:result} provides the experimental results and analysis. Finally, we conclude our work in Section \ref{sec:conclusion}.

\section{Preliminaries}\label{sec:preliminaries}

\subsection{Ransomware Detection: Prior Work}

The generic characteristic of the crypto-ransomware family is to maliciously search and encrypt users' files and provide decryption key only in exchange of a ransom. Very often crypto-ransomware performs targeted attack on specific extensions, e.g., {\fontfamily{pcr}\selectfont .doc, .jpg, .pdf}, or the directories that are more probable to contain important user data, e.g., {\fontfamily{pcr}\selectfont My Documents} folder in {\fontfamily{pcr}\selectfont Windows OS}. Prevalent locker/crypto-ransomware detection techniques heavily rely on inspecting program execution, monitoring system/API calls, signature/template matching in program control flow, and data monitoring for critical file modification. For example, Kharraz et al.\cite{kharraz2015cutting,kharraz2016unveil} analyzed various ransomware families and showed that the majority of them implement naive locking or encryption techniques. Therefore, they proposed a scheme called UNVEIL to detect ransomware infection by using automatically generated artificial user environments where these artificial environments are created for any suspicious activities and constantly monitored. However, this approach is not suitable for lightweight application since deployment and monitoring of artificial user puts significant overhead on execution time and resources without considering kernel applications. Andronio et al. \cite{andronio2015heldroid} proposed an android-based locker-ransomware detection scheme, called HelDroid, using common ransomware characteristics, such as functions to lock screen for android devices. This approach is very much platform-oriented and may not be extendable to commodity processing units for accurate detection. Scaife et al. \cite{scaife2016cryptolock} proposed an early-warning scheme by analyzing variable type changes, similarity measurements, and entropy of user data. This technique, by nature, requires significant data analysis and is not suitable for early detection. Sgandurra et al. \cite{sgandurra2016automated} presented a dynamic analysis of ransomware for higher detection accuracy using signature matching and monitoring dominant features such as API calls. However, such dominant features can be hidden via obfuscated sub-routines in the malware. Moussaileb et al. \cite{moussaileb2018ransomware} presented a machine learning-based ransomware detection technique by monitoring file system traversal using decoy folders, with the observation that the majority of ransomware start their encryption process from the root of the hard disk. However, this technique is prone to high false decision in case any goodware mimics the ransomware's traversal behavior or show no traversal signature at all.

As we see, these techniques utilize software-centric features to distinguish ransomware from goodware. In addition to partial coverage, these approaches also suffer from different challenges, e.g., program and memory overhead for creating virtual users and environments \cite{kharraz2016unveil}, latency and computational overhead for user storage data analysis \cite{spisak2016hardware, sgandurra2016automated}, etc. In contrast, our proposed technique solely relies on hardware-level micro-architectural information that remains unaffected in case of program obfuscation, stealthy execution, and infection strength, and is readily available in modern processors requiring zero hardware overhead.

\subsection{Micro-architectural Event Monitoring for Malware Detection} \label{sec:hpc_for_security}

Hardware-based micro-architectural event monitoring offers a fine-grain filtering for individual executions, can collect multi-dimensional information, and provides a faster data collection with respect to the software-only monitoring schemes. For example, one or more HPCs in the PMU can sample how many times a pre-defined event (enabled by the associated architecture), such as cache misses, occurs during the program runtime to evaluate the performance of the system under test. PMUs in ARM and Intel x86 architectures can be accessed and controlled via lightweight software modules such as {\fontfamily{pcr}\selectfont likwid} and {\fontfamily{pcr}\selectfont perf} tools \cite{likwidgithub,perfgithub}.

Although, the primary motivation behind having performance monitors in hardware was to aid software developers by providing real-time feedback to diagnose bugs or identify bottlenecks in the software for the given the hardware platform, such hardware insights can aid for malware and anomaly detection as well. However, micro-architectural characteristics of both goodware and malware programs can be noisy due to the diffusion of multiple program executions within a given time window. Therefore, it is extremely difficult to characterize and distinguish a malicious program just by simple observation of execution traces. Tang et al. \cite{tang2014unsupervised} proposed anomaly-based malware detection using HPC data. Wang et al. \cite{wang2015confirm} proposed a low-cost validation tool, namely ConFirm, to detect the malicious modifications in the firmware of embedded systems by creating internal check-points to monitor HPC data. Malone et al. \cite{malone2011hardware} analyzed static and dynamic program modifications to detect malicious firmware by modeling the architectural characteristics of benign programs using linear regression. Most of these approaches considered various classes of malware and rootkits into one class of malicious program and, therefore, combined generic signature traces without emphasizing the intrinsic nature (and subsequent micro-architectural activities) of the malware itself. As a result, these techniques require a large amount of data for analysis, produces relatively large false decisions, and are not suitable for early detection schemes focusing specific families of malware, such as crypto-ransomware. 

Very recently, Alam et al. \cite{alam2019ratafia} presented a scheme, named Ratafia, that leveraged HPC-provided micro-architectural event traces for runtime detection of ransomware. This technique utilized the fast-Fourier transform of event traces to identify prominent features for generating a one-class classifier and designed a watchdog program with LSTM-based autoencoders for anomaly detection in program execution trace. However, the technique was developed and validated only for a handful of ransomware variants and do not provide any scalability information for such.

Our proposed technique, on the other hand, is extensively analyzed for a large number of ransomware and goodware for different scenario and provides successful detection with low false decisions. Our technique requires only 2ms of execution traces for correctly classifying a ransomware versus a goodware.

\subsection{Security Enhancement via Advanced Machine Learning Techniques} \label{sec:ml_for_security}
One major obstacle for HPC-assisted malware detection is that the same micro-architectural event can occur in a similar manner (i.e., frequency count and event profile) during a benign (valid) operation. Therefore, it leads to possible false detection. However, carefully constructed machine-learning (ML) techniques can learn and differentiate such events to identify anomaly with a higher confidence \cite{demme2013feasibility}. Two fundamental requirements for deploying hardware-assisted anti-malware techniques are: (1) selecting high-fidelity micro-architectural features and events, and (2) choosing efficient machine learning techniques for classification. 

In this work, we collect micro-architectural event traces from selected HPCs in a timeseries fashion. We use the recurrent neural network with long short-term memory (LSTM) architecture \cite{hochreiter1997long, gers2002learning, alam2019ratafia}. LSTM is widely popular in timeseries analysis, especially in audio-visual domain, since it helps maintaining a constant error for recurrent networks and continues to learn over multiple timestamps in a series fashion. The decisions that the LSTM generates depend on the current input, previous output, and previous memory of the LSTM cell iteslf. The LSTM cells are usually outside the general flow of the recurrent network and can filter the signals they receive with their own set of weights. Alam et al. \cite{alam2019ratafia} used LSTM-based auto-encoder for anomaly detection in event traces. In contrast, our work implements a RNN-based binary classifier where a LSTM network is followed by Global Average Pooling (GAP) Layers to reduce any overfitting in the model. Details of the adopted ML architecture is discussed in Section \ref{sec:methodology}.2.

\section{RanStop: A Hardware-assisted Runtime Crypto-ransomware Detector} \label{sec:methodology}

In this section, we describe our proposed technique, RanStop, for the runtime detection of crypto-ransomware via micro-architectural event monitoring using HPCs. We have built our framework utilizing the key observations presented in Demme et al. \cite{demme2013feasibility}:

\begin{enumerate}
	
	\item 
	The semantics of a program (goodware or ransomware) do not change significantly over different variants of similar functionality and class.
	 
	\item 
	While accomplishing a particular task (benign or malicious), there exist subtasks that cannot be radically modified and should exhibit similar micro-architectural footprints.

\end{enumerate}

\begin{figure}[!t] 
	\begin{center}
		\includegraphics[keepaspectratio,width=0.85\columnwidth, height=7.8in]{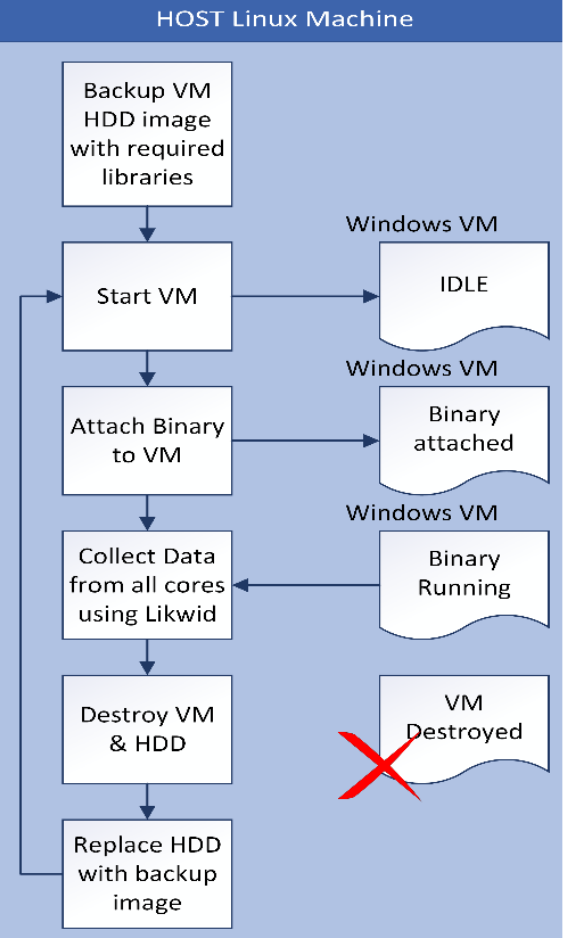}
		\caption[]{HPC data collection scheme for ransomware and goodware.} \label{fig:DataCollection}
	\end{center}
\end{figure}

\begin{figure*}[!htbp]
	\begin{center}
		\includegraphics[width=.8\textwidth]{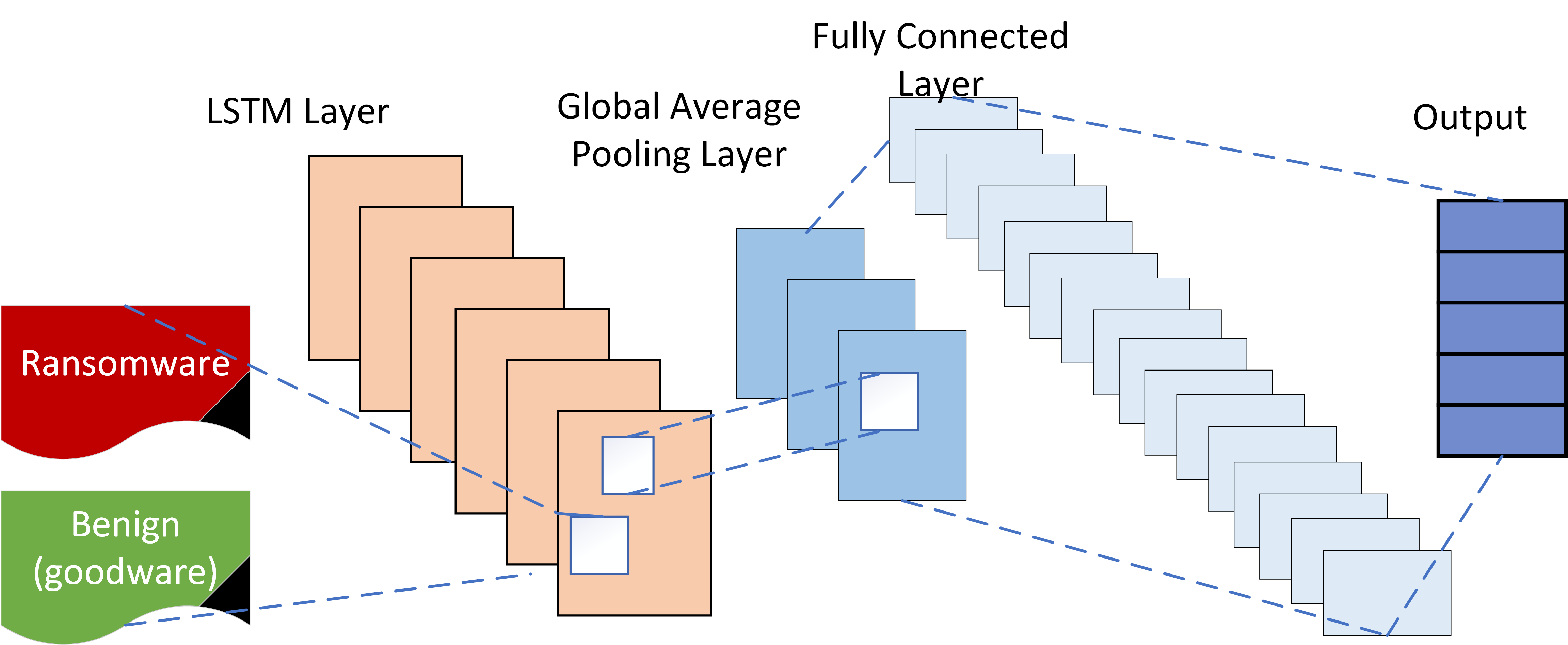}
        \vspace{0.1in}
		\caption[]{LSTM-based model generation for crypto-ransomware detection.} \label{fig:NeuralModel}
	\end{center}
\end{figure*}

This observation exhibits the potential of building a hardware-assisted crypto-ransomware detector for early recognition by analyzing micro-architectural events at runtime. Even though the micro-architectural information collected via HPCs are noisy; it can enable identifying (or, separating, at least) crypto-ransomware from the benign activities if chosen correctly with proper optimization. This is possible because there exists a significant amount of similar semantic characteristics among multiple variants of crypto-ransomware due to the similarity of their attack behavior \cite{moussaileb2018ransomware}. HPCs, being oblivious to the undergoing program, are able to collect this multi-dimensional signature that can be put under further scrutiny via machine learning for ransomware versus goodware identification.

Figure \ref{fig:BlockDiagram} shows the high-level workflow of our porposed RanStop technique. It consists of following major steps: (1) program database creation; (2) micro-architectural event monitoring and data collection in a timeseries fashion; (3) LSTM-based predictive model generation (training); and, finally, (4) testing, validation, and runtime detection. We discuss the details of each step in the following subsections.

\subsection{Program Database Creation} 
The very first step of RanStop is creating a database of benign (goodware) programs and publicly available variants of crypto-ransomware, as shown in Figure \ref{fig:BlockDiagram}. This program database is fed to RanStop framework for subsequent micro-architectural data collection. The platform, source, and size of the program database is discussed in details in Sections \ref{sec:result}.1-2.

We note that each ransomware executable was manually tested to make sure it did not throw any runtime error; i.e., the ransomeware executables were compatible with the execution environment and had access to necessary resources similar to any real-life infection. Similarly, the benign program database were also tested in a similar fashion so the the acquired HPC dataset were not corrupted due incompatibility.

One key point for program database creation is that the goodware database should contain different families of benign programs with various workload. Especially, one should also consider computationally intensive programs (e.g., disk encryption programs such as VeraCrypt \cite{veracrypt2018}) that perform legit but similar operations with respect to that of a crypto-ransomware. The motivation behind is to offer similar semantic characteristics to different sub-routines of the crypto-ransomware as well as generic user specific benign programs. Additionally, the non-encryption benign binaries provide resemblance to silent crypto-ransomware which does not start execution at the very first moment of infection but resort to stealthy operation in the background to other legit programs.

The idea behind choosing ransomware and goodware with similar characteristics is to make sure that the proposed framework is capable of identifying even the smallest differences and is not over/under-fitted due to noise \cite{witten2016data}. For example, as experimented in Alam et al. \cite{alam2019ratafia}, a one-class classifier trained with random benign programs may tend to separate crypto-ransomware more accurately from a text editor program; but may not distinguish from a disc encryption or file zipping program. Therefore, we adopt a well-balanced database with significant number and variants of ransomware and goodware. This well-balanced training scheme allows to reduce false positive and false negative by the classifier. Note that the RanStop framework is readily scalable to a larger dataset, as we discuss in Section \ref{sec:result}, and it allows the user to re-train (update) the initial model for finer detection with emerging threats.

\subsection{Micro-architectural Event Monitoring and Data Collection}

In this work, we have developed our framework for an experimental {\fontfamily{pcr}\selectfont Linux OS} setup, whereas a majority of the real-life crypto-ransomware are designed for {\fontfamily{pcr}\selectfont Windows OS}. Therefore, we execute all programs (both ransomware and goodware) in a {\fontfamily{pcr}\selectfont Windows OS virtual machine} hosted by the experimental {\fontfamily{pcr}\selectfont Linux} system. (For detailed configuration of the experimental platform, please see Section \ref{sec:result}.1.) This approach is taken because -- (1) it averts the risk of cryto-ransomware encrypting the collected HPC data which is stored in a separate administrative-privileged directory; (2) the Linux system provided inhospitable environment in case any ransomware binary manages to escape the Windows VM, so that the rest of the networked systems (if any) in the experimental setup is unaffected.

The crypto-ransomware and benign programs from the database are executed in a random fashion to monitor and collect micro-architectural information from the processor using HPCs. We use the open-source tool {\fontfamily{pcr}\selectfont likwid} \cite{likwidgithub} for capturing hardware data from embedded HPCs. To make sure that the virtual machine offers the same workload signature with or without infection, we collected the HPC data in the following fashion, as shown in Figure \ref{fig:DataCollection}:
\begin{itemize}
    \item The Windows VM is hosted and run with a complete library of programs to replicate a real-life workload.
    \item The program under test (ransomware or goodware) is pinned to run inside the VM with no thread/resource limitations.
    \item \si{Likwid} is used to collect and store timestamp data from all the CPU cores in the \si{Linux} host machine.
    \item Once the targeted timeseries data is collected (e.g., by completion of the program or timeout); The VM is destroyed along with its virtual storage completely wiped to reduce any residual noise.
    \item A new VM replaces the old one (e.g., corrupted one, if infected by ransomware while collecting ransomware data) with a backup storage image having the same state as prior to running the program.
    \item Multiple iterations are performed to collect all possible micro-architectural events with randomized execution order, so that there exists no systematic data and memory correlation, irrespective to ransomware or goodware execution.
    
\end{itemize}

The collected micro-architectural events (and respective values) are then categorized into different performance groups on the basis of event context, such as {\fontfamily{pcr}\selectfont BRANCH, L2\_DATA, ICACHE}, etc., along with associated event counts and metrics, as shown in Table \ref{tab:eventgroup}. These associated events are given input to the next stage LSTM network for ML-based predictive model building. More details on these hardware features with appropriate optimizers and event counts are discussed in Section \ref{sec:methodology}.4 and Section \ref{sec:result}.3.

\begin{center}
\begin{table}[!htbp]
	\small
	\centering
	\caption{Performance Event Groups and Associated Metrics for Data Collection via Likwid.}
	\label{tab:eventgroup}
    \vspace{-0.1in}
	\begin{tabular}{|c|c|l|}
		\hline
		\textbf{No} & \textbf{Group Name} & \multicolumn{1}{c|}{\textbf{Metrics}} \\ \hline
		\multirow{4}{*}{1} & \multirow{4}{*}{BRANCH} & Branch rate \\ \cline{3-3} 
		&  & Branch misprediction rate \\ \cline{3-3} 
		&  & Branch misprediction ratio \\ \cline{3-3} 
		&  & Instructions per branch \\ \hline
		\multirow{1}{*}{2} & \multirow{1}{*}{CLOCK} & Uncore Clock {[}MHz{]} \\ \hline
		\multirow{4}{*}{3} & \multirow{4}{*}{\begin{tabular}[c]{@{}c@{}}CYCLE\\ ACTIVITY\end{tabular}} & Cycles without execution {[}\%{]} \\ \cline{3-3} 
		&  & Cycles with stalls due to L1D {[}\%{]} \\ \cline{3-3} 
		&  & Cycles with stalls due to L2 {[}\%{]} \\ \cline{3-3} 
		&  & \begin{tabular}[c]{@{}l@{}}Cycles w/o execution due to memory {[}\%{]}\end{tabular} \\ \hline
		4 & DATA & Load to store ratio \\ \hline
		\multirow{5}{*}{\textbf{5}} & \multirow{5}{*}{FLOPS\_DP} & DP MFLOP/s \\ \cline{3-3} 
		&  & AVX DP MFLOP/s \\ \cline{3-3} 
		&  & Packed MUOPS/s \\ \cline{3-3} 
		&  & Scalar MUOPS/s \\ \cline{3-3} 
		&  & Vectorization ratio \\ \hline
		\multirow{4}{*}{6} & \multirow{5}{*}{ICACHE} & L1I request rate \\ \cline{3-3} 
		&  & L1I miss rate \\ \cline{3-3} 
		&  & L1I miss ratio \\ \cline{3-3} 
		&  & L1I stall rate \\ \hline
		\multirow{6}{*}{7} & \multirow{6}{*}{L2\_DATA} & L2D load bandwidth {[}MBytes/s{]} \\ \cline{3-3} 
		&  & L2D load data volume {[}GBytes{]} \\ \cline{3-3} 
		&  & L2D evict bandwidth {[}MBytes/s{]} \\ \cline{3-3} 
		&  & L2D evict data volume {[}GBytes{]} \\ \cline{3-3} 
		&  & L2 bandwidth {[}MBytes/s{]} \\ \cline{3-3} 
		&  & L2 data volume {[}GBytes{]} \\ \hline
		\multirow{3}{*}{8} & \multirow{3}{*}{L2\_CACHE} & L2 request rate \\ \cline{3-3} 
		&  & L2 miss rate \\ \cline{3-3} 
		&  & L2 miss ratio \\ \hline
		\multirow{6}{*}{9} & \multirow{6}{*}{L3\_DATA:} & L3 load bandwidth {[}MBytes/s{]} \\ \cline{3-3} 
		&  & L3 load data volume {[}GBytes{]} \\ \cline{3-3} 
		&  & L3 evict bandwidth {[}MBytes/s{]} \\ \cline{3-3} 
		&  & L3 evict data volume {[}GBytes{]} \\ \cline{3-3} 
		&  & L3 bandwidth {[}MBytes/s{]} \\ \cline{3-3} 
		&  & L3 data volume {[}GBytes{]} \\ \hline
		\multirow{3}{*}{10} & \multirow{3}{*}{L3\_CACHE} & L3 request rate \\ \cline{3-3} 
		&  & L3 miss rate \\ \cline{3-3} 
		&  & L3 miss ratio \\ \hline
		\multirow{6}{*}{11} & \multirow{6}{*}{TLB\_DATA} & L1 DTLB load misses \\ \cline{3-3} 
		&  & L1 DTLB load miss rate \\ \cline{3-3} 
		&  & L1 DTLB load miss duration {[}Cyc{]} \\ \cline{3-3} 
		&  & L1 DTLB store misses \\ \cline{3-3} 
		&  & L1 DTLB store miss rate \\ \cline{3-3} 
		&  & L1 DTLB store miss duration {[}Cyc{]} \\ \hline
		\multirow{3}{*}{12} & \multirow{3}{*}{TLB\_INSTR} & L1 ITLB misses \\ \cline{3-3} 
		&  & L1 ITLB miss rate \\ \cline{3-3} 
		&  & L1 ITLB miss duration {[}Cyc{]} \\ \hline
		\multirow{3}{*}{13} & \multirow{3}{*}{UOPS} & Issued UOPs \\ \cline{3-3} 
		&  & Executed UOPs \\ \cline{3-3} 
		&  & Retired UOPs \\ \hline
		\multirow{3}{*}{14} & \multirow{3}{*}{UOPS\_EXEC} & Used cycles ratio {[}\%{]} \\ \cline{3-3} 
		&  & Unused cycles ratio {[}\%{]} \\ \cline{3-3} 
		&  & Avg stall duration {[}cycles{]} \\ \hline
		\multirow{3}{*}{15} & \multirow{3}{*}{UOPS\_ISSUE} & Used cycles ratio {[}\%{]} \\ \cline{3-3} 
		&  & Unused cycles ratio {[}\%{]} \\ \cline{3-3} 
		&  & Avg stall duration {[}cycles{]} \\ \hline
		\multirow{3}{*}{16} & \multirow{3}{*}{UOPS\_RETIRE} & Used cycles ratio {[}\%{]} \\ \cline{3-3} 
		&  & Unused cycles ratio {[}\%{]} \\ \cline{3-3} 
		&  & Avg stall duration {[}cycles{]} \\ \hline

\end{tabular}
\end{table}
\end{center}

\subsection{LSTM-based Predictive Model Generation (training and validation)}

For generating a ML-based predictive model, that will be used for detecting ransomware via runtime event monitoring, we utilize LSTM-based recurrent neural network \cite{hochreiter1997long,gers2002learning} for performing a timeseries classification of the micro-architectural event signatures. As shown in Figure \ref{fig:NeuralModel}, the collected hardware event values are provided as the input features to the LSTM layer of the ML architecture. The reason for using this specific neural model is already discussed in Section \ref{sec:preliminaries}.3. A standard LSTM cell can remember values over time interval making it best candidate to classifying timeseries data of micro-architectural events from the process execution.

After the LSTM layer, a global average pooling (GPA) layer is used to reduce the intrinsic training features created by the LSTM layer. This can significantly improve the model accuracy and prevent over-fitting of the tensor. For our specific application, it reduces the spatial dimension of a three dimension model to one dimension as necessary for the following layer. For example, a tensor of dimension \si{a \times b \times c} is reduced to dimension of \si{1 \times 1 \times d} after the GAP layer. The produced features at this stage is then fed as input to the fully connected layer of the neural network architecture for generating the binary classification model (i.e., goodware versus ransomware).

\subsection{Model Generation, Validation, and Detection}

The final steps of the proposed RanStop is to generation of the predictive model based on the given dataset, and validation and deployment of the model for runtime detection. It should be noted that the impacts of different optimizers and the micro-architectural performance groups collected in previous steps are not all same for detecting potential crypto-ransomware with high accuracy (see Section \ref{sec:result}). Also, since the number of hardware performance counters are limited on any system, the real time detection program (watchdog) can only be trained to work for certain performance groups and may not swap between monitors to monitor different set of data too often.

\section{Experimental Results} \label{sec:result}

\subsection{Experimental Platform}
The system used as the experimental platform was based on {\fontfamily{pcr}\selectfont Intel Xeon CPU-E3-1225 Coffeelake} processor with a maximum operating frequency of \SI{3.30}{\giga\hertz} and 32GB RAM. It was a single socket quad-core processor with L1, L2, and L3 {\fontfamily{pcr}\selectfont Cache}  of \si{32\kilo B}, \si{256\kilo B}, and \si{8\mega B}, respectively. The operating system was {\fontfamily{pcr}\selectfont Ubuntu 16.04 LTS} and {\fontfamily{pcr}\selectfont likwid-4.3.2} \cite{likwidgithub} was used for HPC event collection.

\subsection{Program Database Creation}
For this work, we considered 80 crypto-ransomware executables and 76 benign (goodware) programs all of which had a execution time of at least 2\si{\milli\second} or more. Irrespective to total runtime, we collected data for the first 2\si{\milli\second} only because our primary objective is an early detection. The crypto-ransomware database was collected from {\fontfamily{pcr}\selectfont VirusShare} \cite{virusshare} which comprised of Windows executable ({\fontfamily{pcr}\selectfont.exe}) files. For the benign programs, we used combination of encryption algorithms provided by the {\fontfamily{pcr}\selectfont OpenSSL} \cite{openssl} and collection of random C programs from {\fontfamily{pcr}\selectfont Github} \cite{awesome-c,ckatas}. The benign encryption binaries are used to encrypt a local directory from {\fontfamily{pcr}\selectfont govdocs1} \cite{govdocs1}.

\subsection{Micro-architectural Event Capture}

\begin{figure*}[!t]
\subfloat{\includegraphics[width = .25\textwidth]{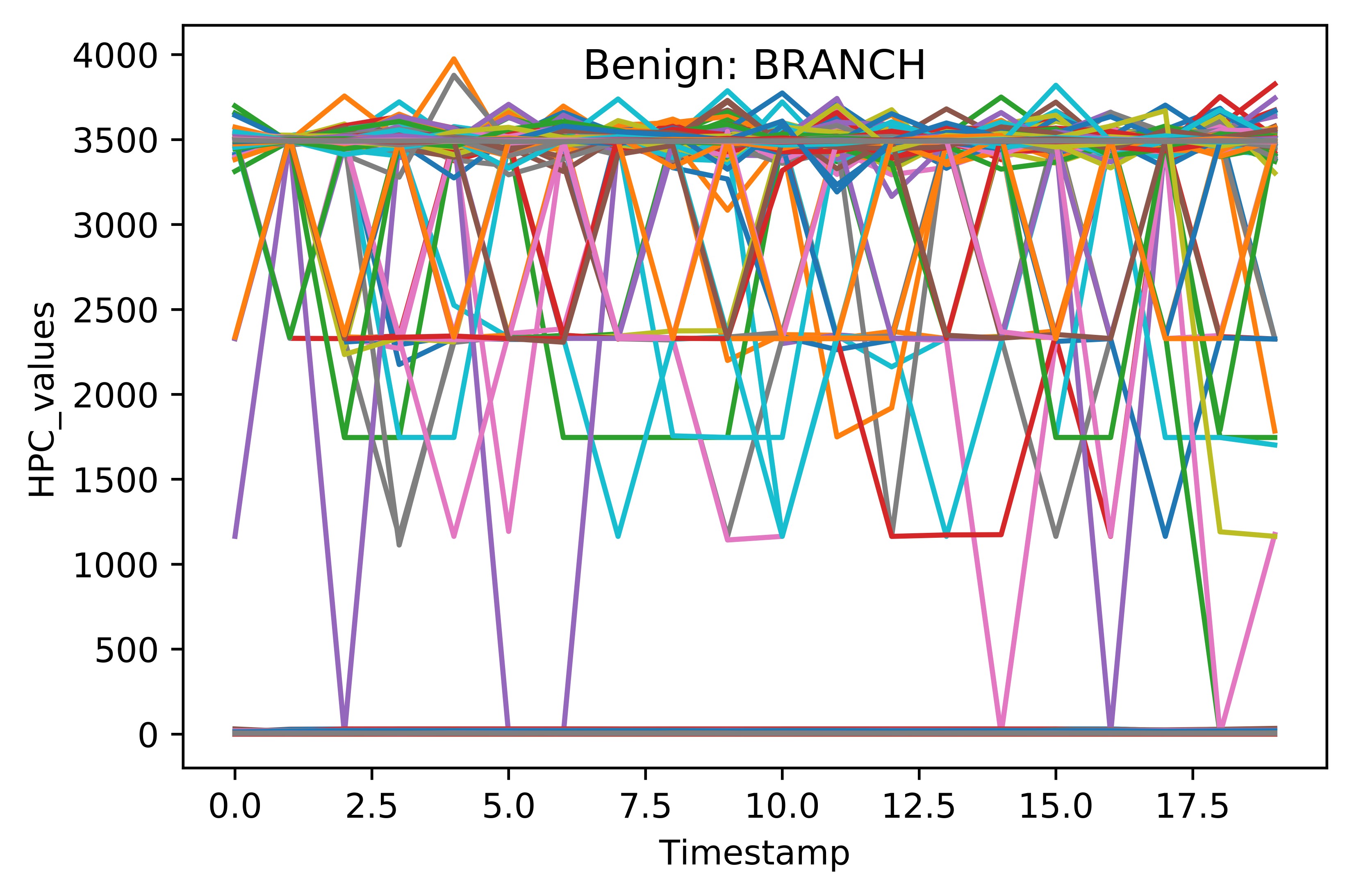}}
\subfloat{\includegraphics[width = .25\textwidth]{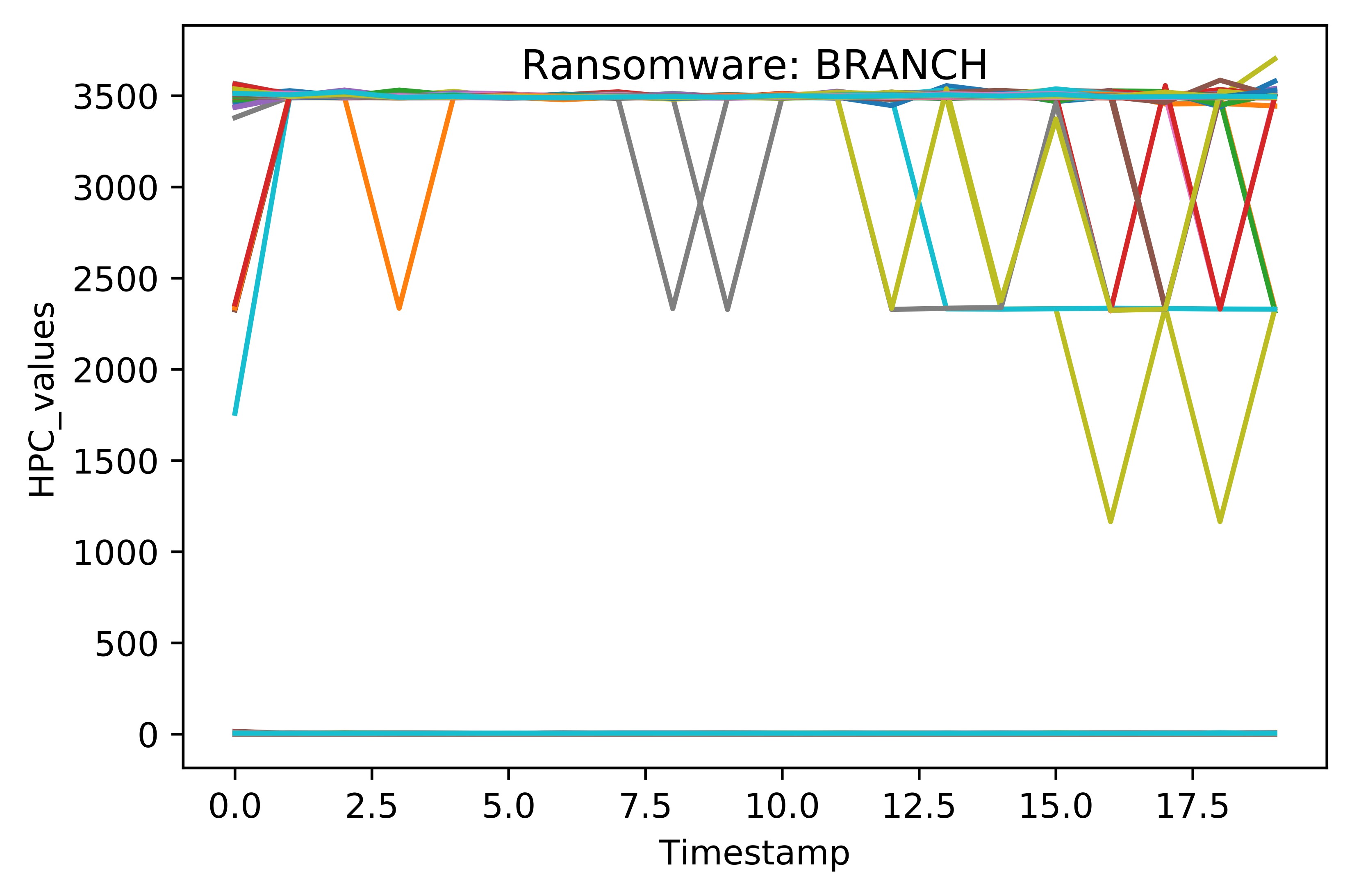}}
\subfloat{\includegraphics[width = .25\textwidth]{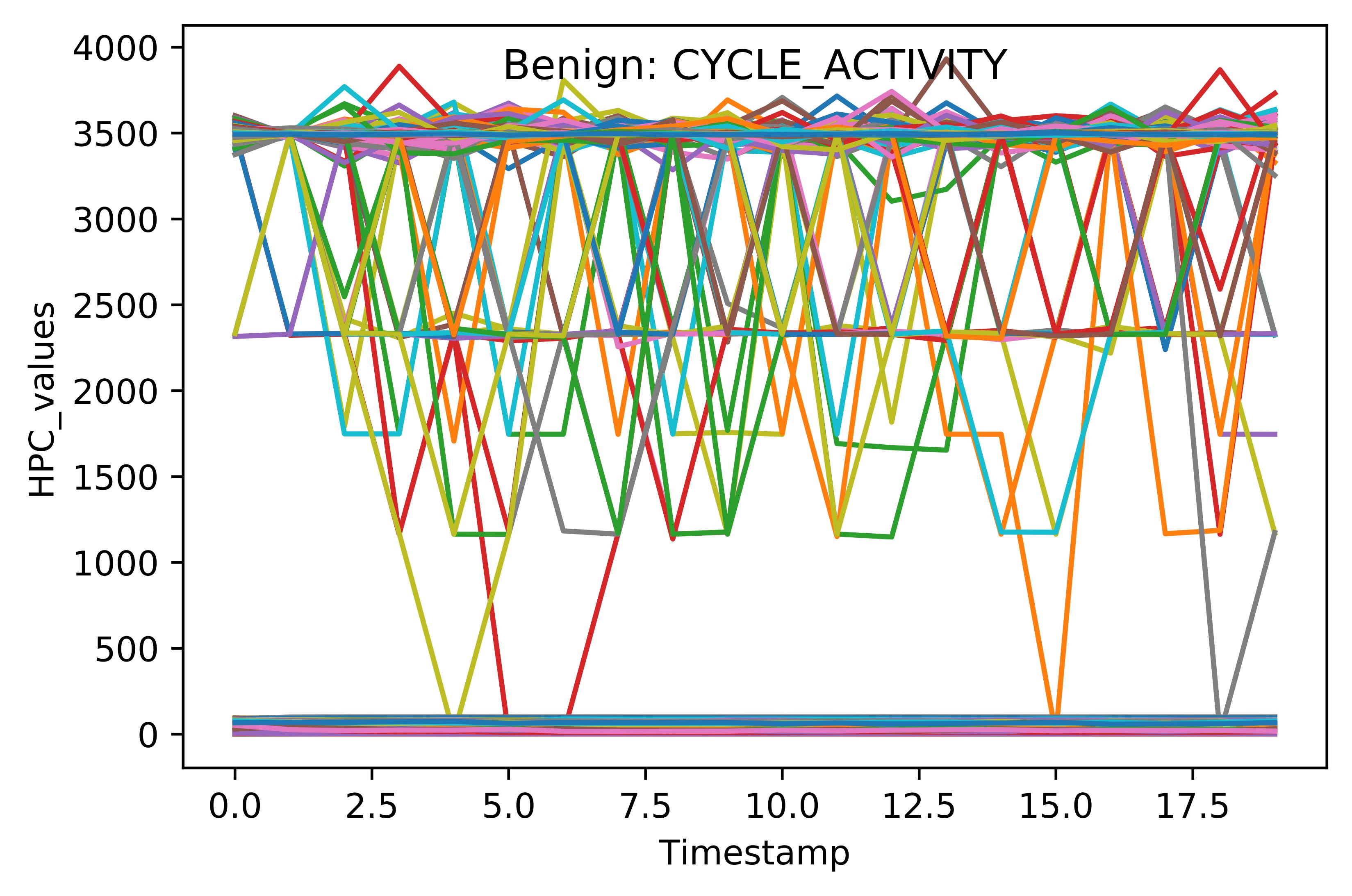}}
\subfloat{\includegraphics[width = .25\textwidth]{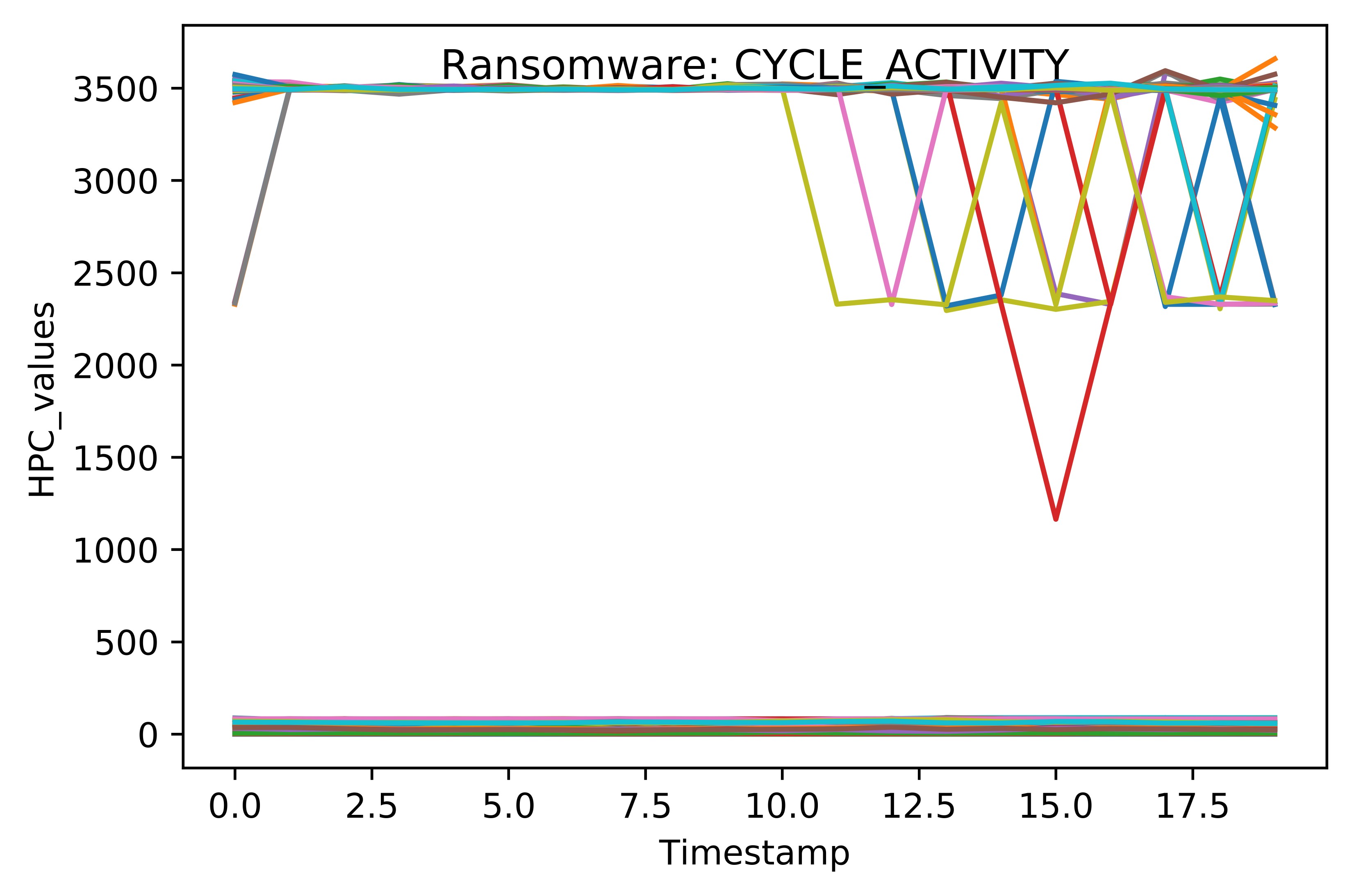}}
\newline
\subfloat{\includegraphics[width = .25\textwidth]{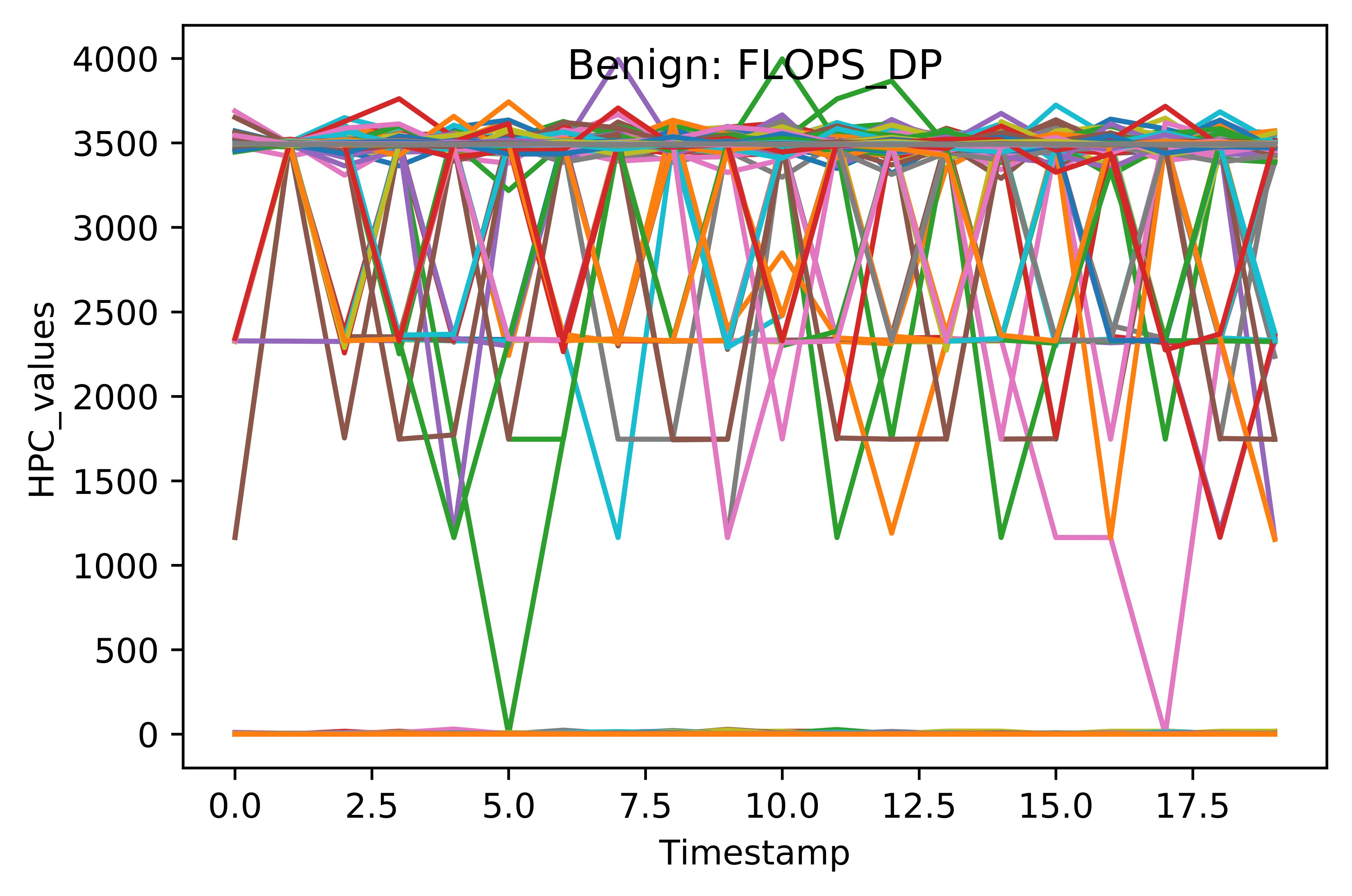}}
\subfloat{\includegraphics[width = .25\textwidth]{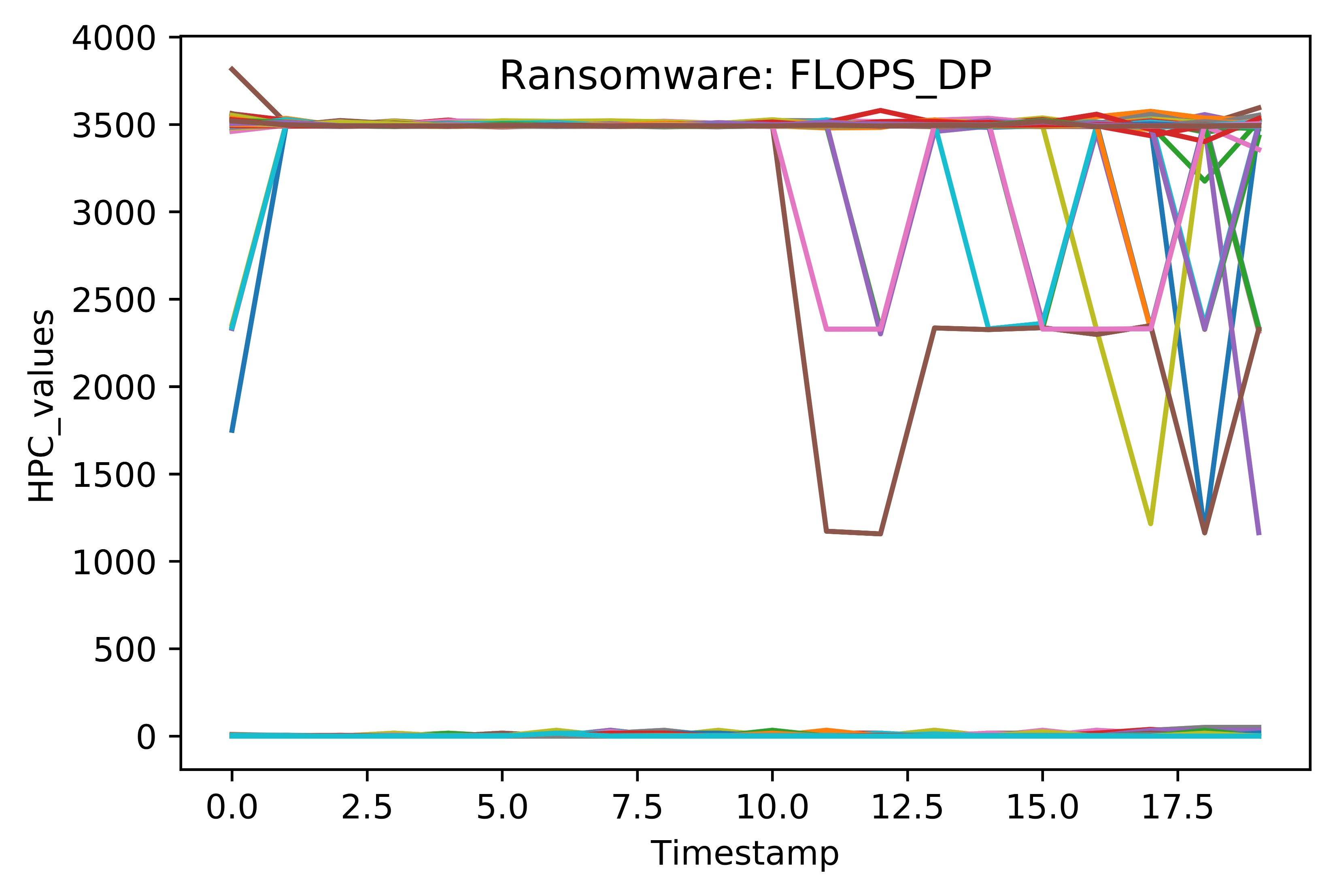}}
\subfloat{\includegraphics[width = .25\textwidth]{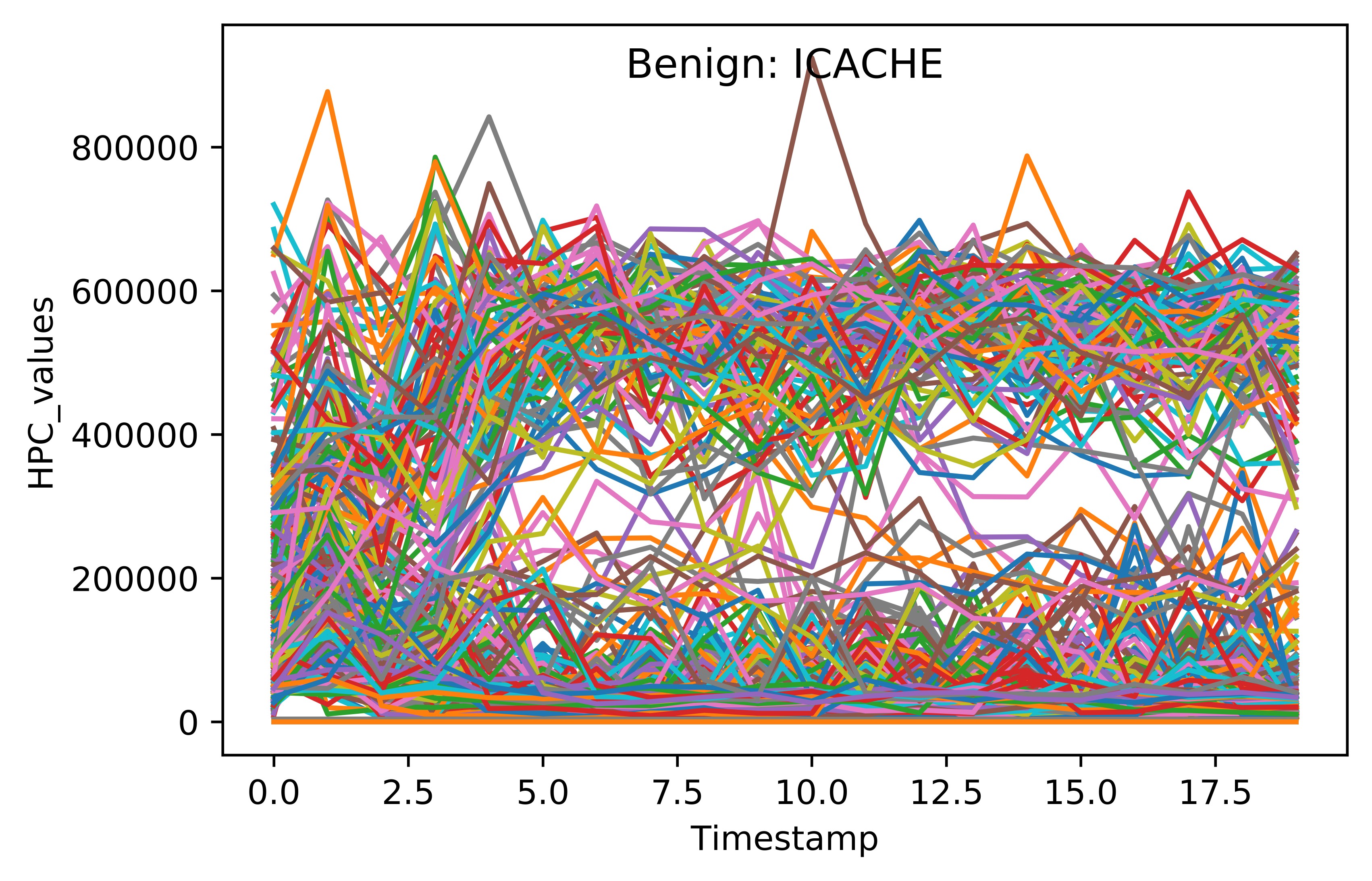}}
\subfloat{\includegraphics[width = .25\textwidth]{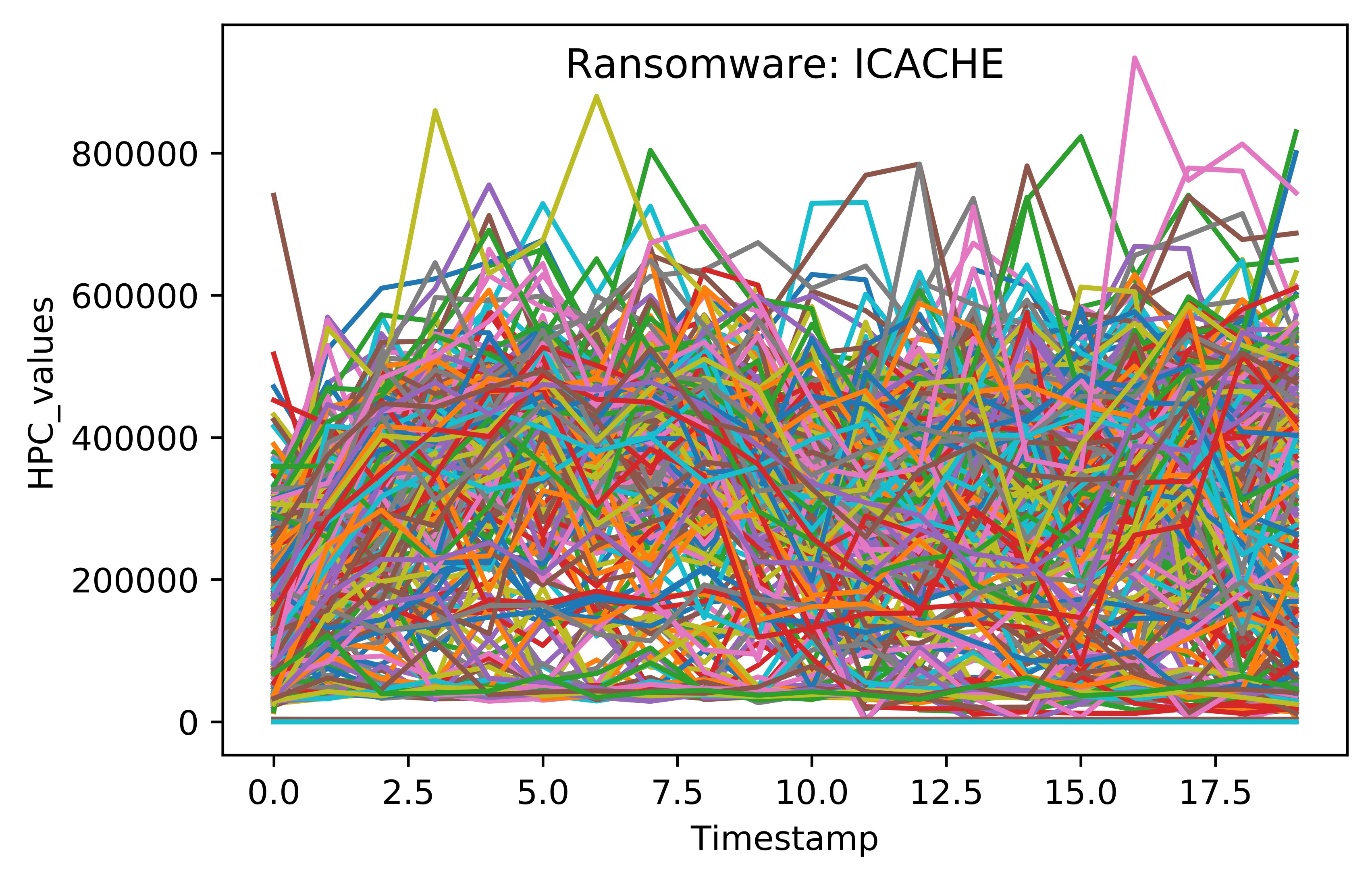}}
\newline
\subfloat{\includegraphics[width = .25\textwidth]{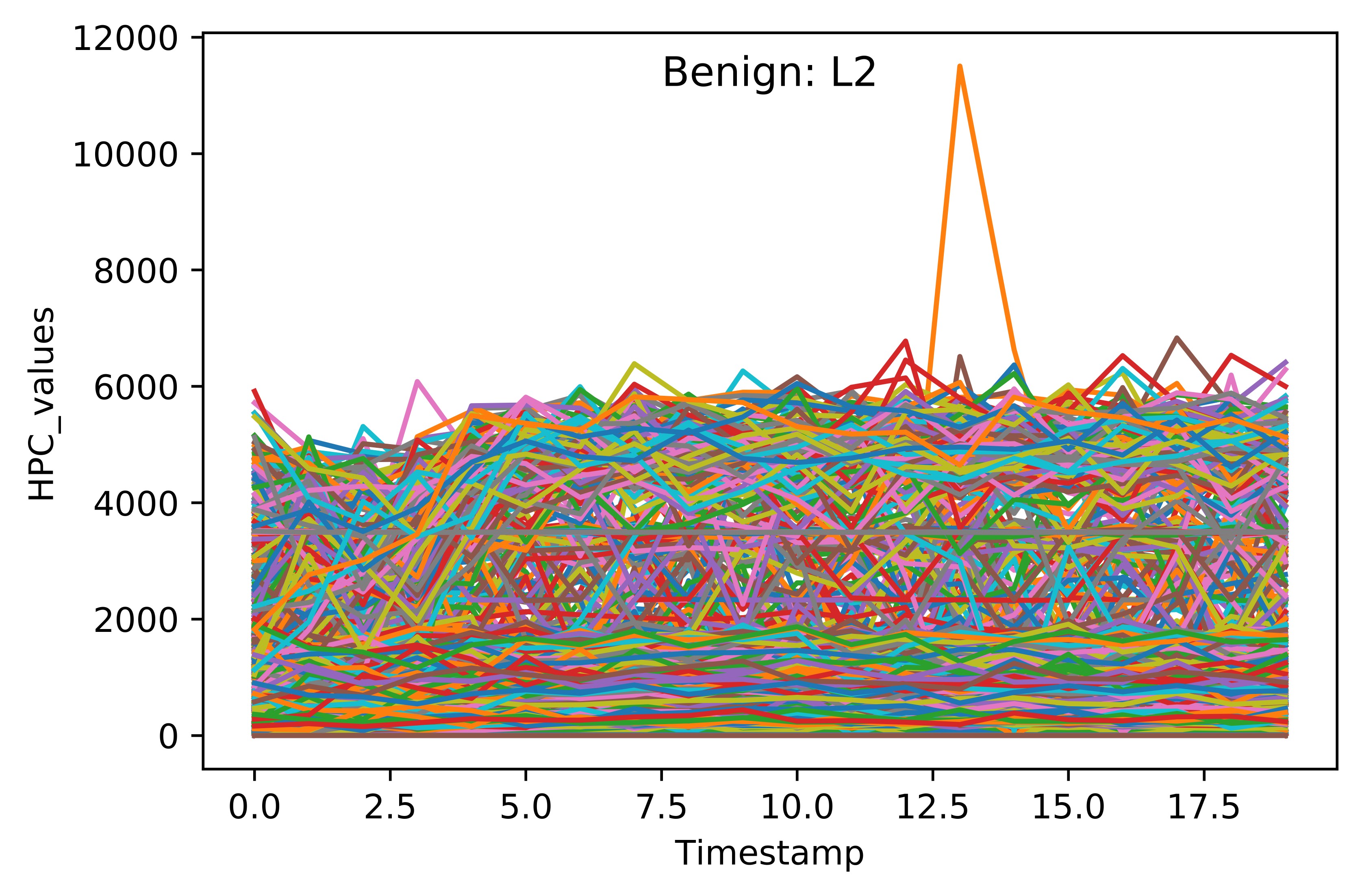}}
\subfloat{\includegraphics[width = .25\textwidth]{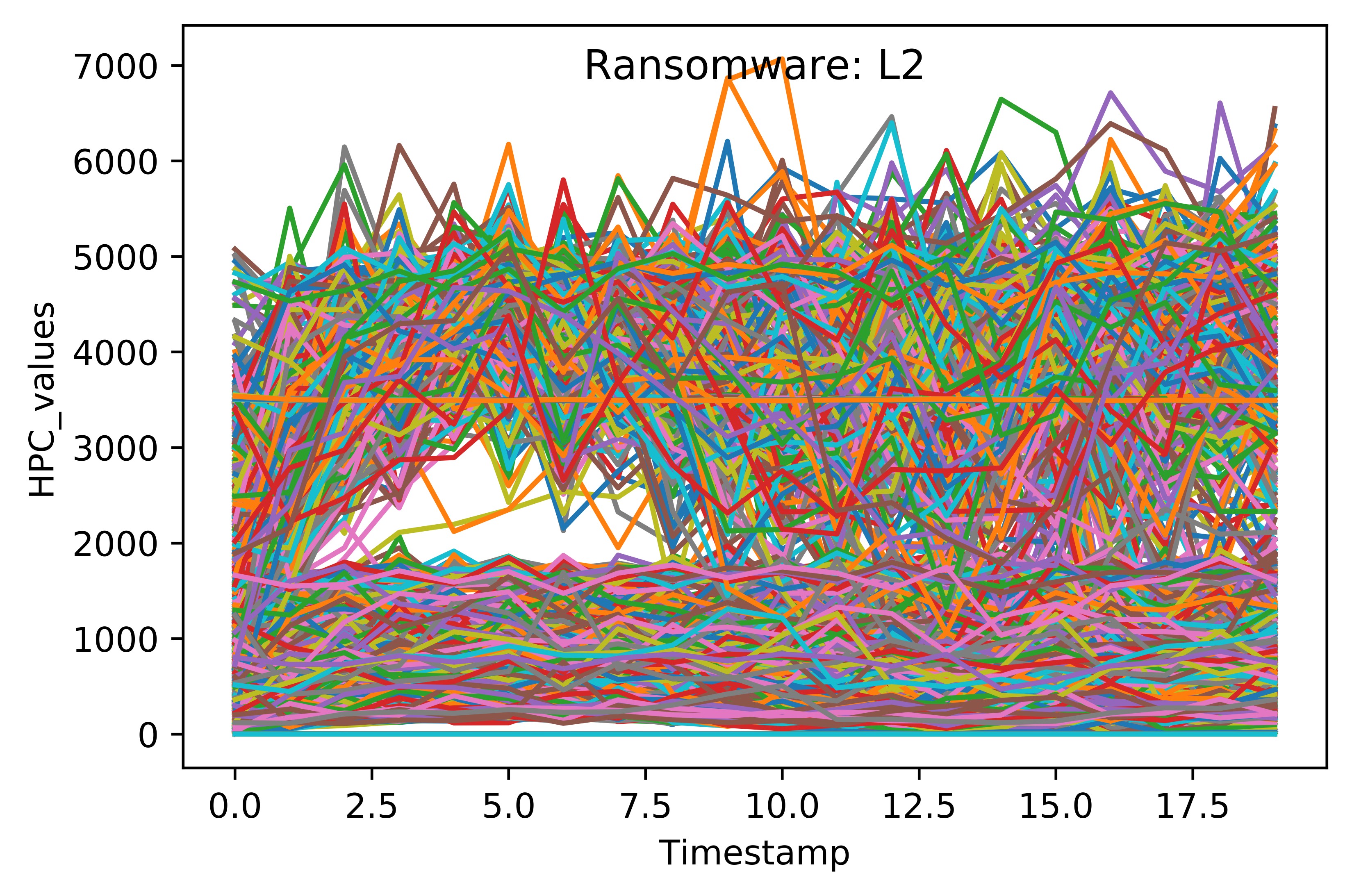}}
\subfloat{\includegraphics[width = .25\textwidth]{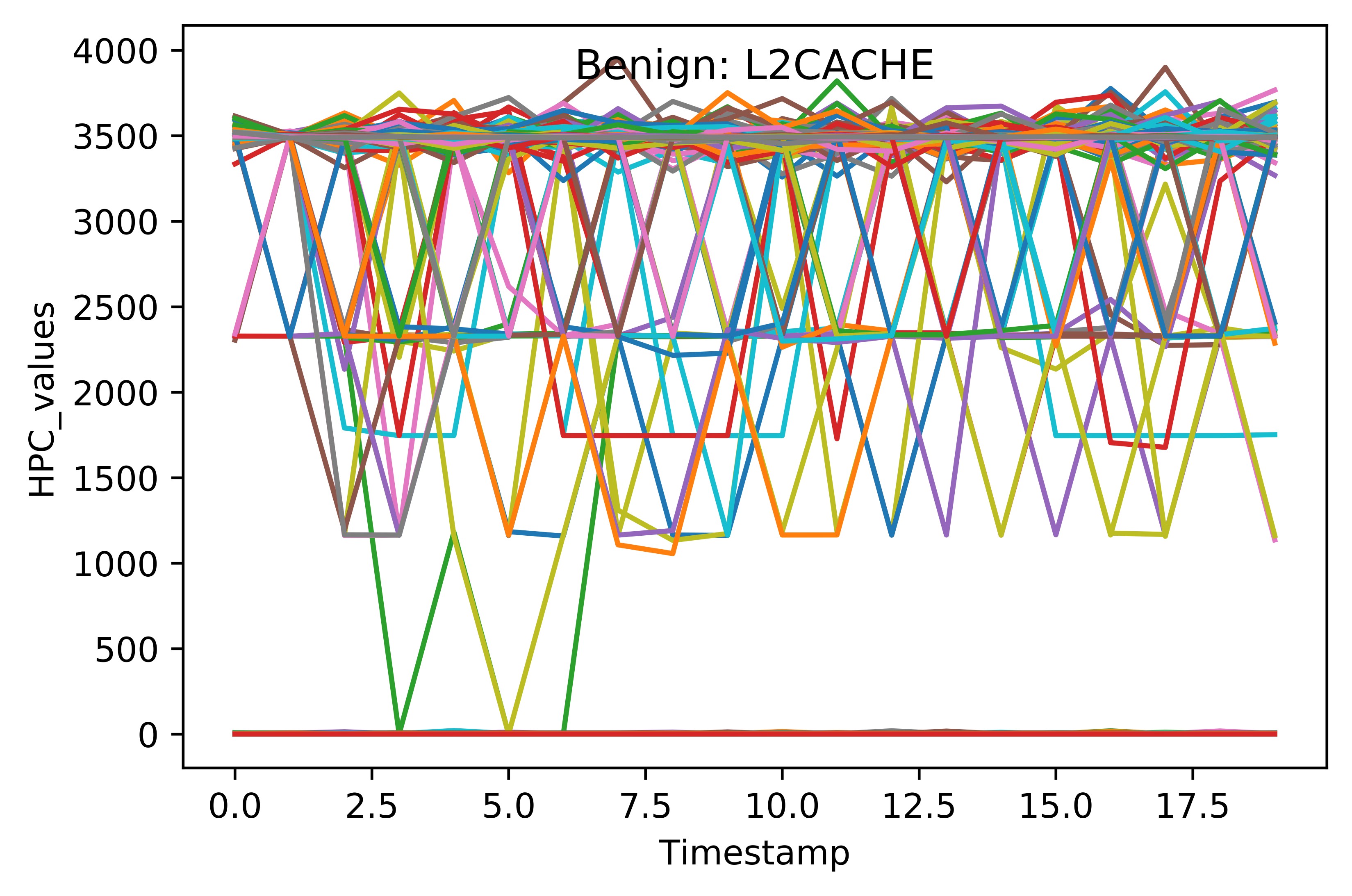}}
\subfloat{\includegraphics[width = .25\textwidth]{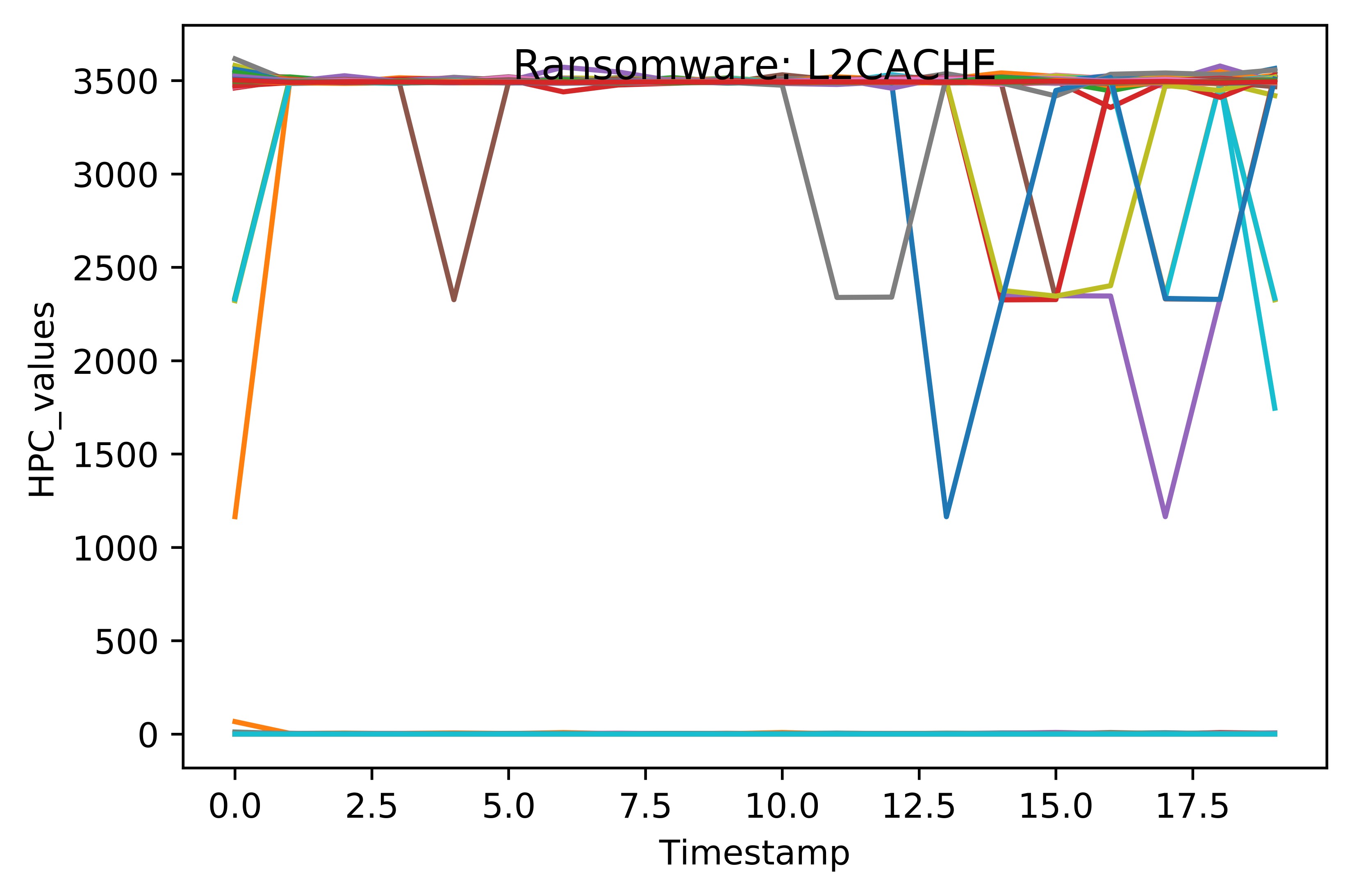}}
\newline
\subfloat{\includegraphics[width = .25\textwidth]{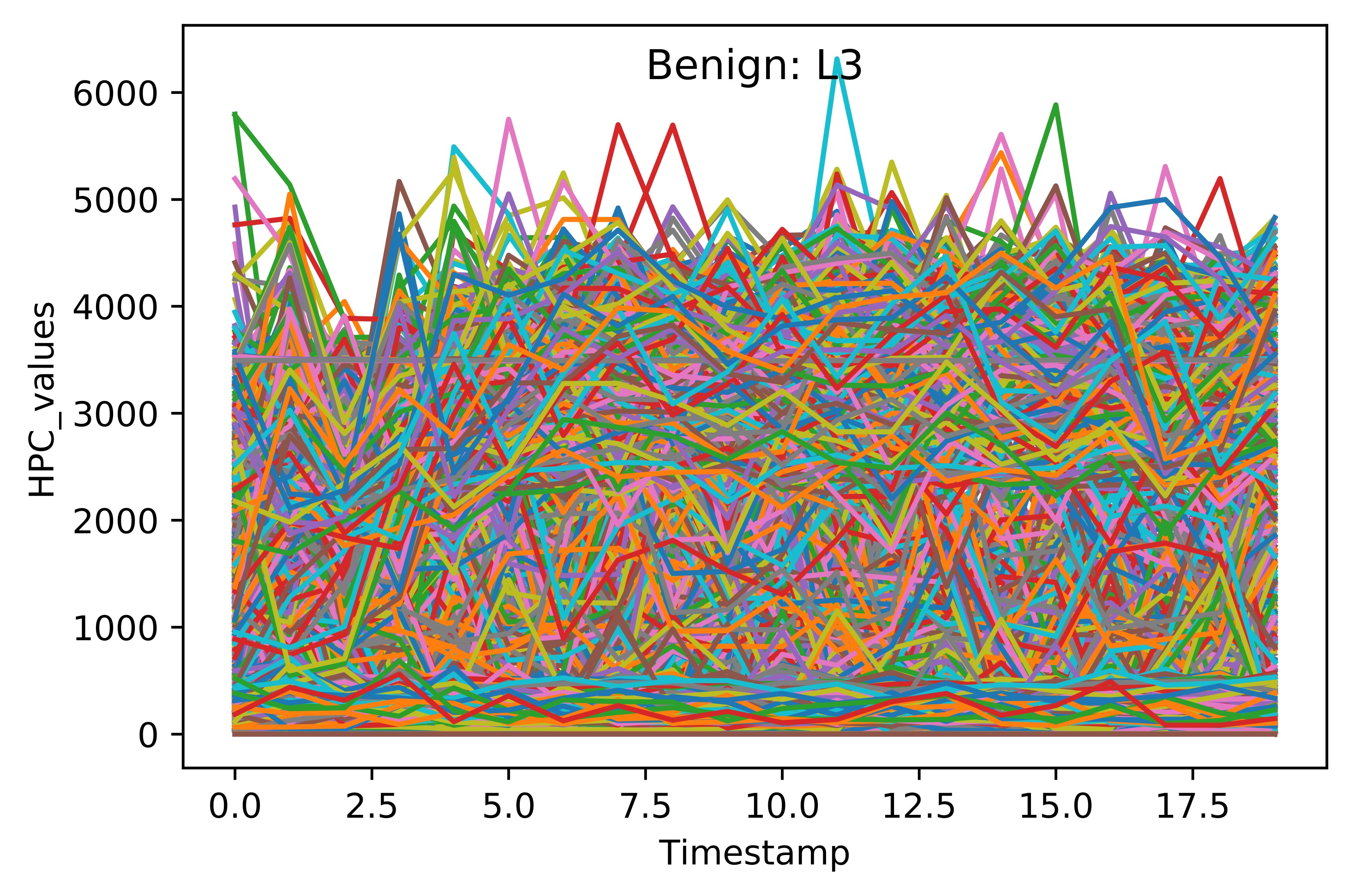}}
\subfloat{\includegraphics[width = .25\textwidth]{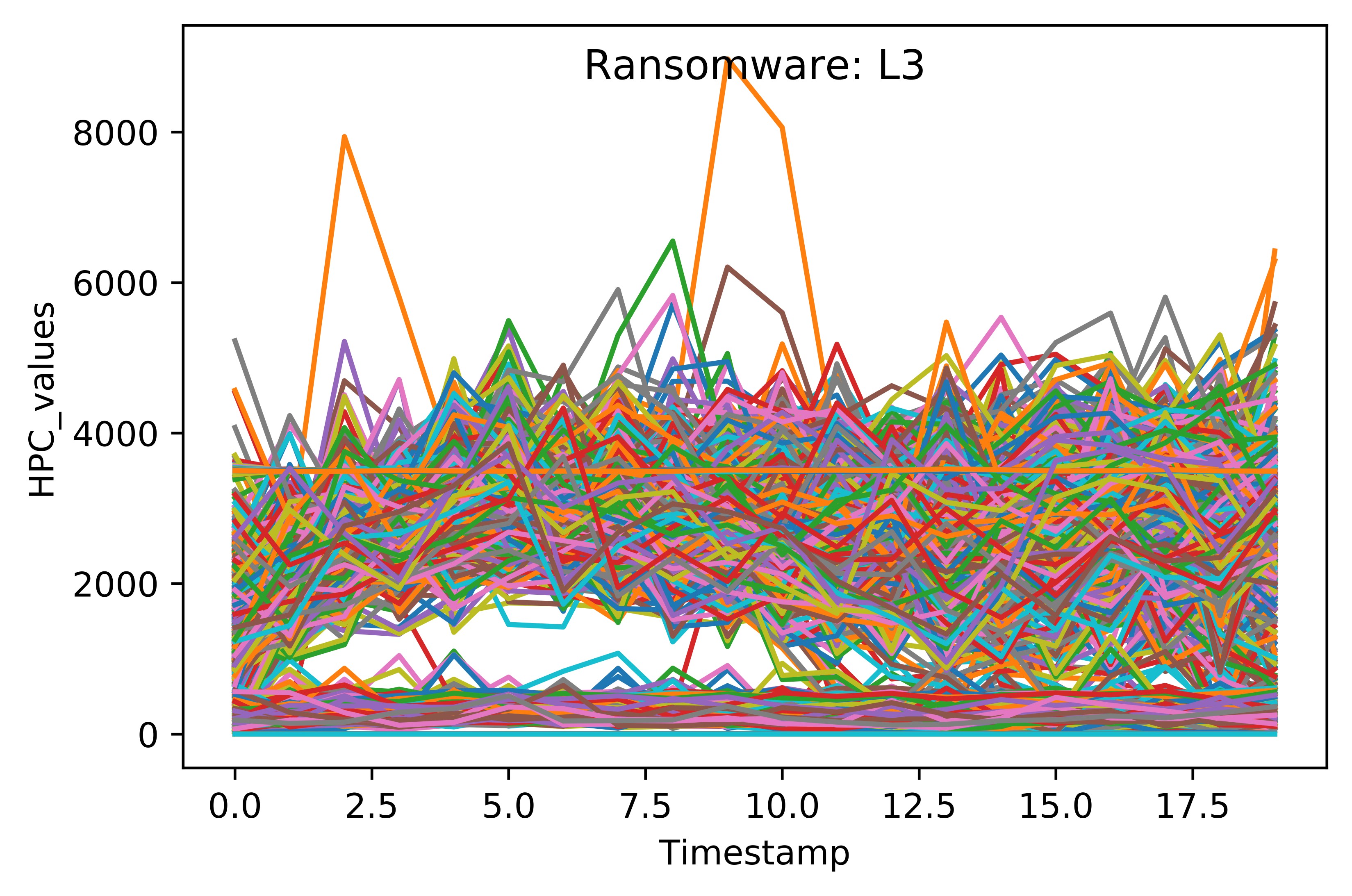}}
\subfloat{\includegraphics[width = .25\textwidth]{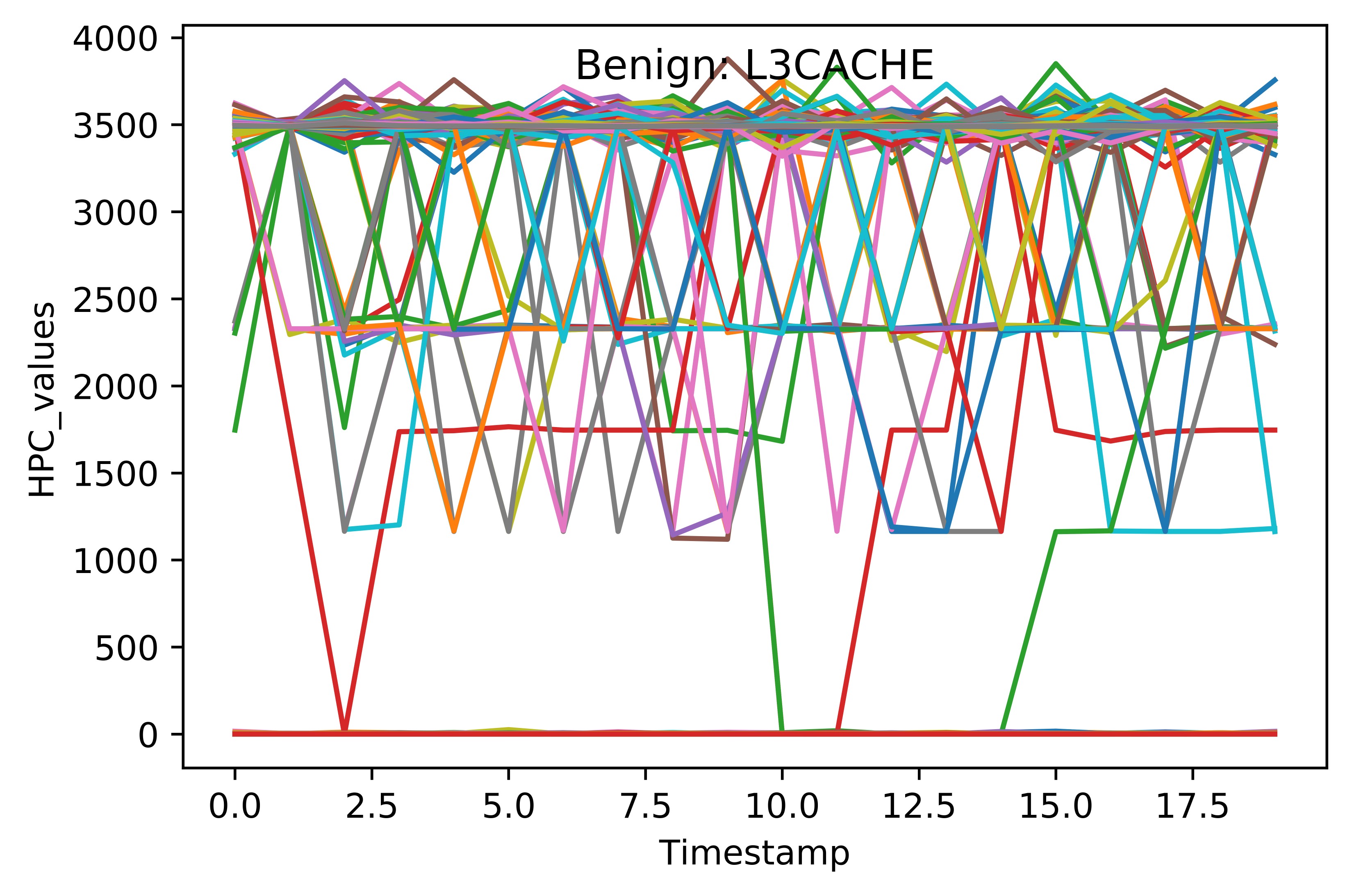}}
\subfloat{\includegraphics[width = .25\textwidth]{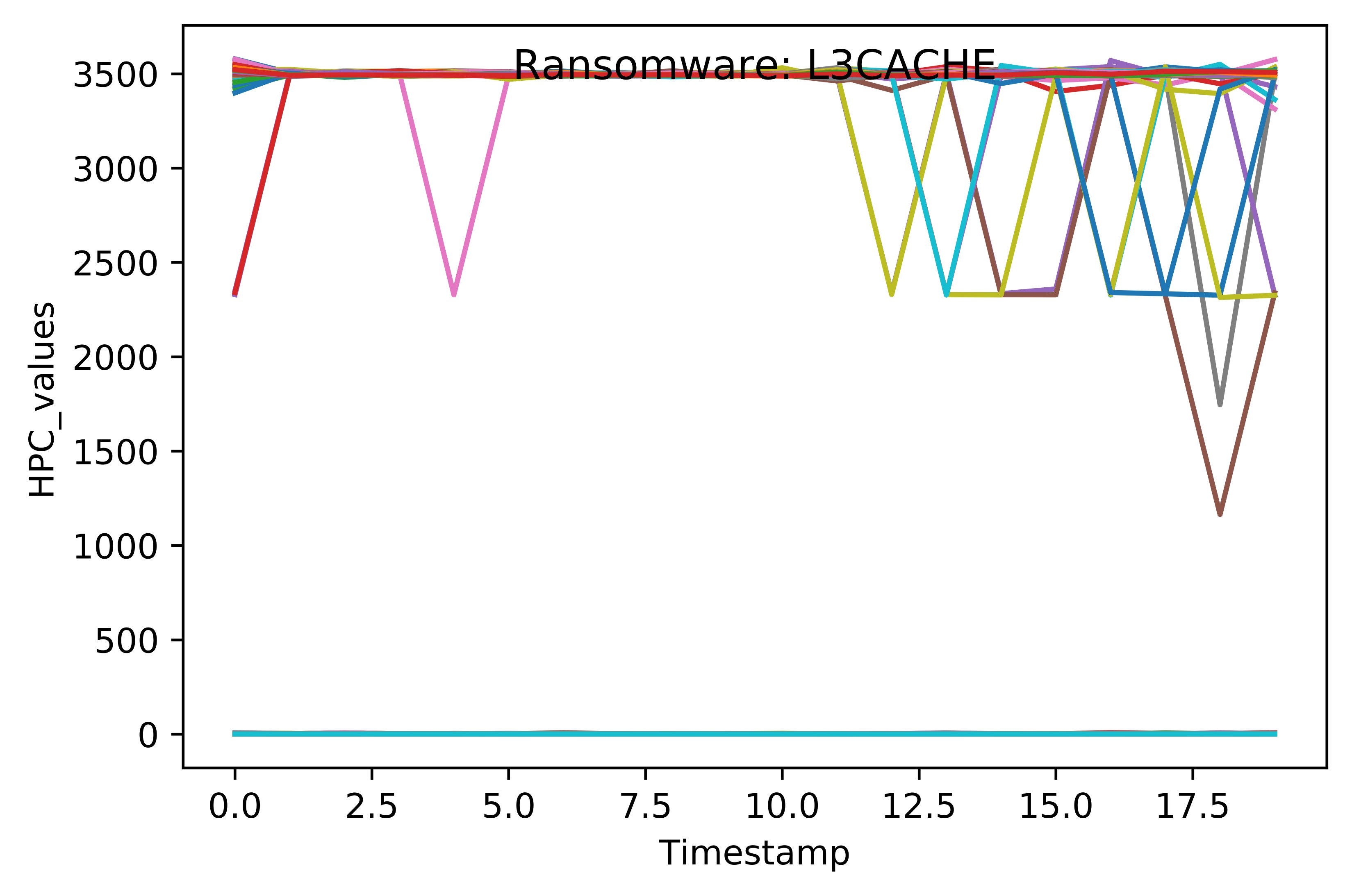}}
\newline
\subfloat{\includegraphics[width = .25\textwidth]{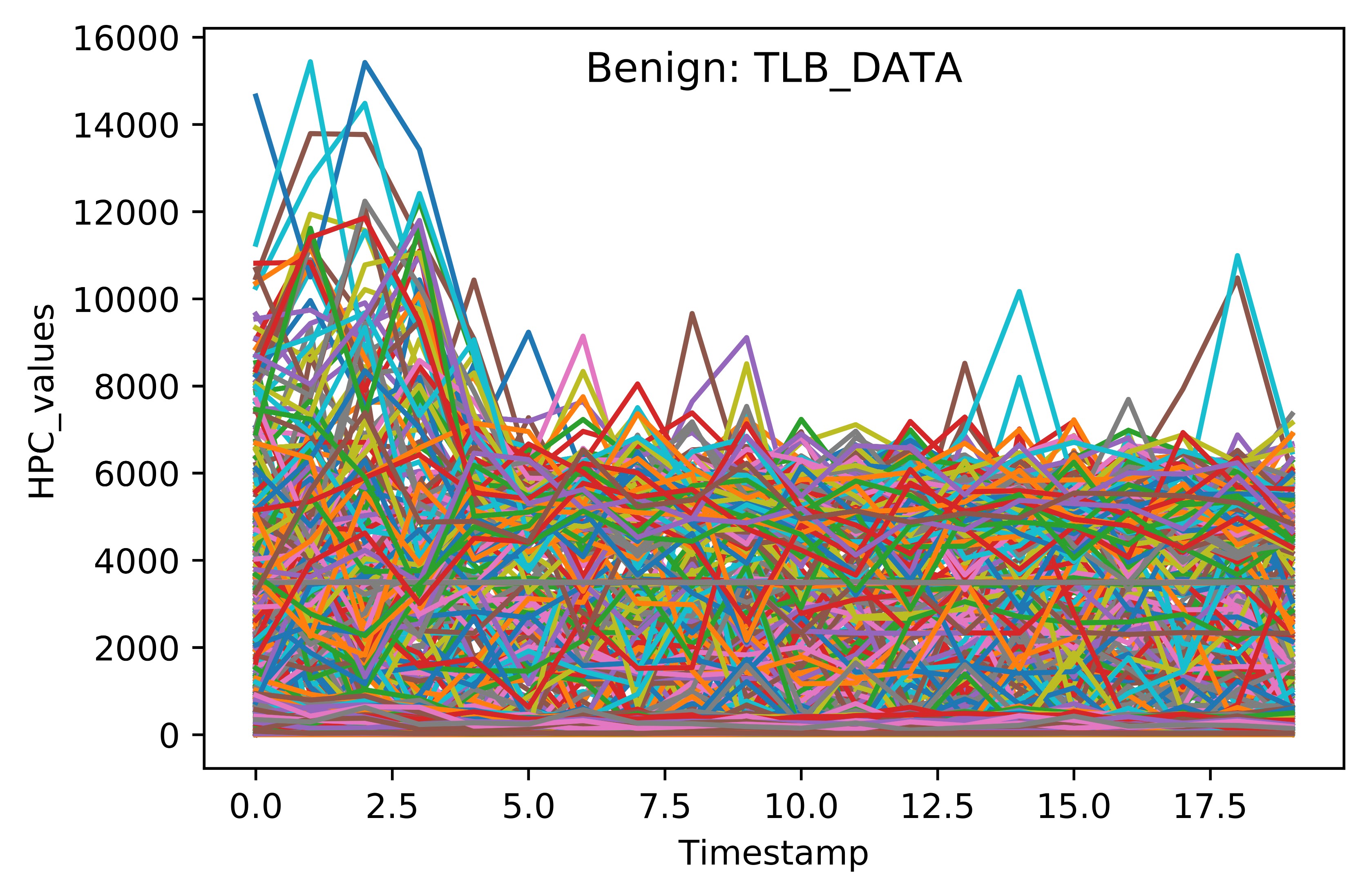}}
\subfloat{\includegraphics[width = .25\textwidth]{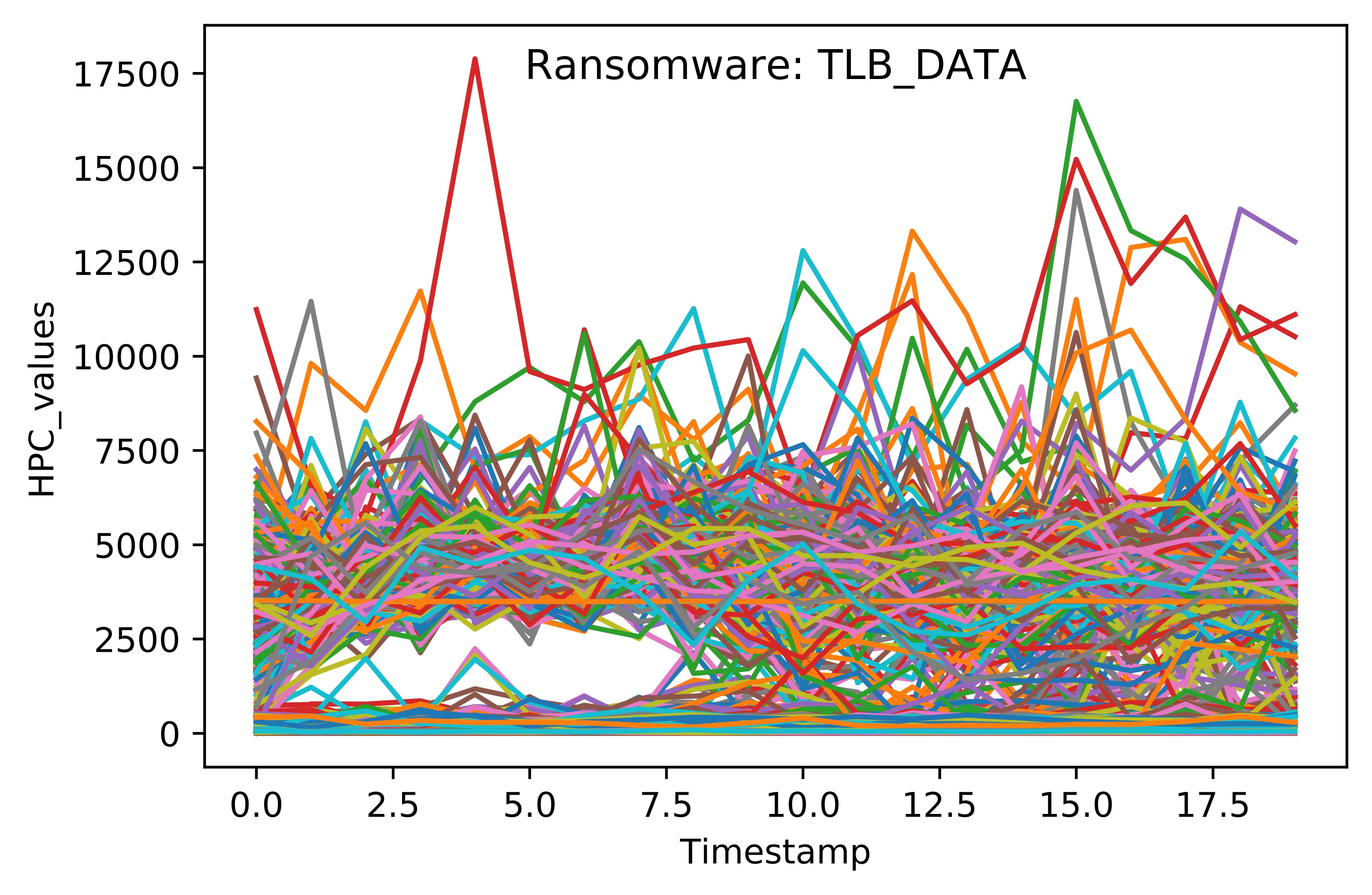}}
\subfloat{\includegraphics[width = .25\textwidth]{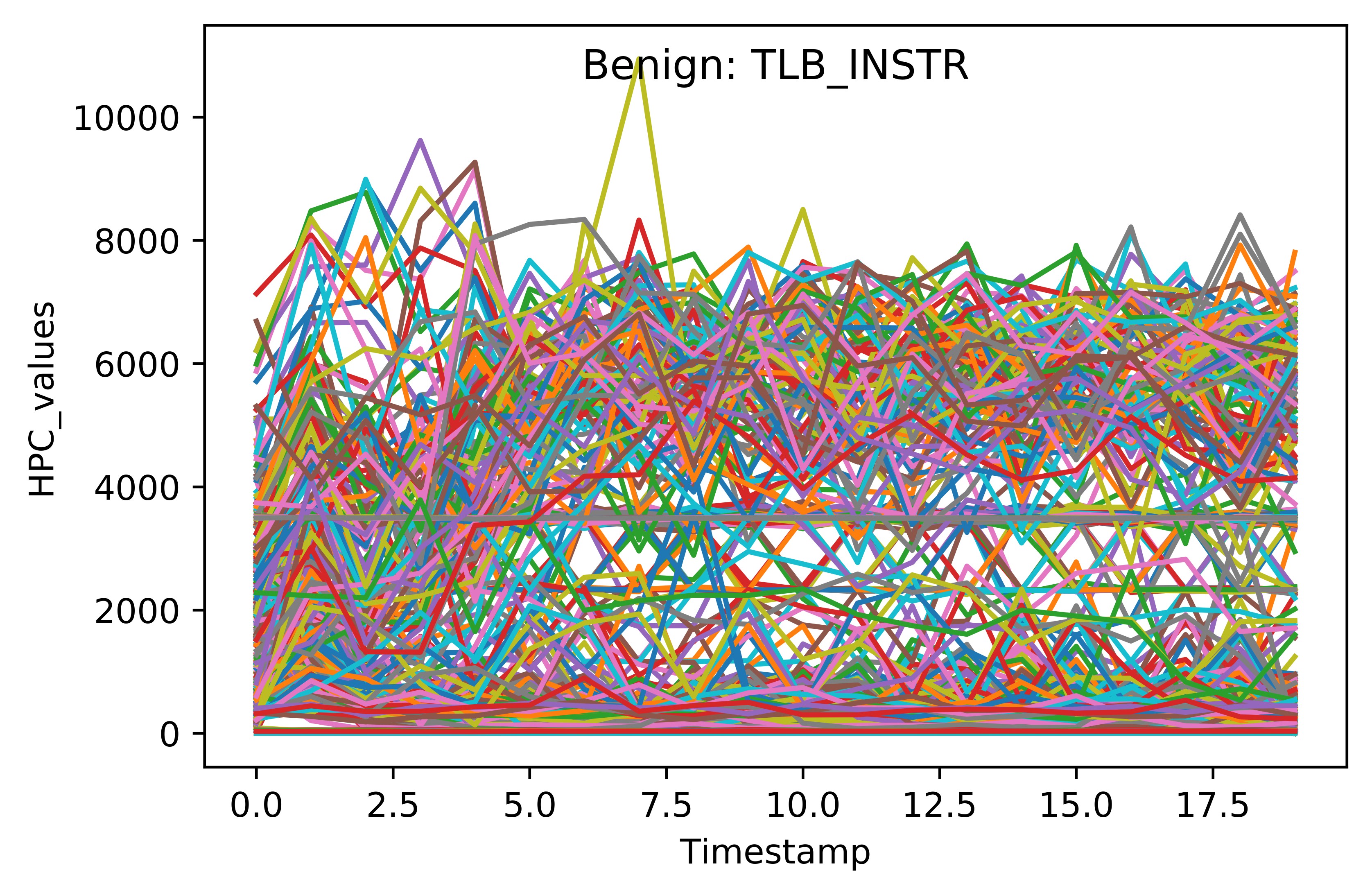}}
\subfloat{\includegraphics[width = .25\textwidth]{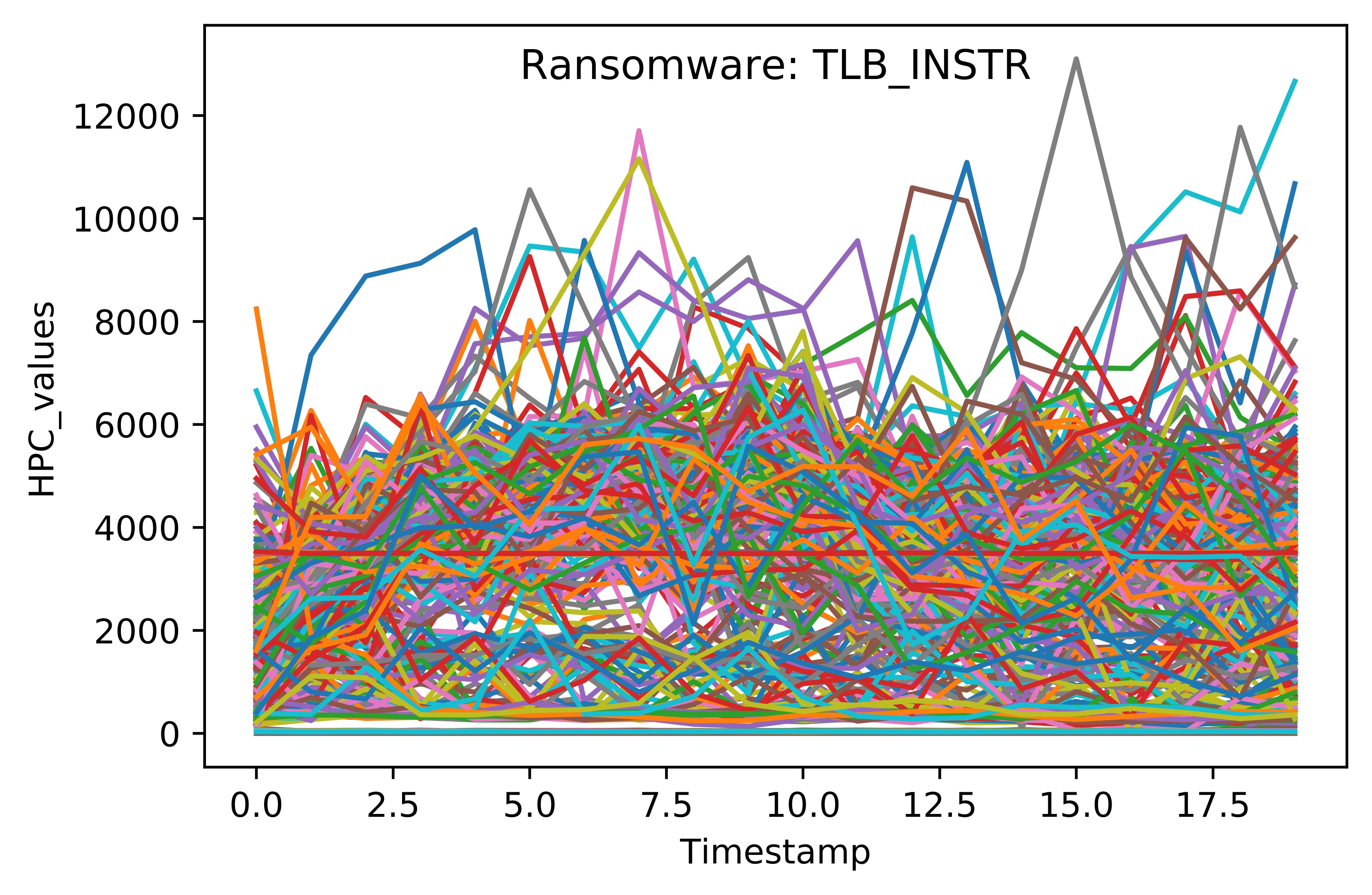}}
\caption{Distribution of example micro-architectural events among benign and ransomwares for different performance groups. X-axis shows 20 timestamps from the start of the execution, each timestamp are 100\si{\micro\second} apart. Y-axis shows respective micro-architectural event count in the embedded hardware performance counter.} 
\label{fig:Distributionbenignransomware}
\end{figure*}

Figure \ref{fig:Distributionbenignransomware} depicts sample hardware events collected for goodware and crypto-ransomware for different performance group. Here, HPC information was collected for 20 timestamps, each being 100\si{\micro\second} apart. Our objective was to accurately detect ransomware within the 2\si{\milli\second} execution threshold; and as we will see in Section \ref{sec:result}.4, we were successful to correctly classify ransomware from a goodware just utilizing hardware information of these 20 timestamps. As one can see, this detection time is too early for a ransomware to make any significant damage, if none.

We also saw that the timeseries data differences between the two classes were not necessarily significantly large to readily distinguish between ransomware versus goodware. Additionally, the differences (or similarities) at some timestamps might have occured due to system noise and additional runtime overhead. Therefore, it was necessary that the developed detection scheme ccould reduce any noise and optimized the intrinsic features to accurately identify ransomware threats.


It should be noted that each performance group can collect \si{4} or fewer micro-architectural events due to hardware limitations. Because, for our experimental system, only four general-purpose HPCs are available in each core when hyperthreading is enabled \cite{intel64manual}. In addition to micro-architectural event count, {\fontfamily{pcr}\selectfont likwid} readily provides scalar information based on different performance metrics as shown in Table \ref{tab:eventgroup}. For the ease of analysis, we consider available pre-processed metric information for feature selection, training, and testing in the subsequent steps, rather than naively using raw hardware event counts which are often noisy and require data pre-processing such as scaling and alignment \cite{tang2014unsupervised,demme2013feasibility}.

\begin{table}[!t]
\small
\centering
\caption{Accuracy with 70\% Training Dataset}
\label{tab:70}
\vspace{-0.1in}
\begin{tabular}{|c|c|c|c|c|}
\hline
\textbf{\begin{tabular}[c]{@{}c@{}}\end{tabular}} & \textbf{Adadelta} & \textbf{Adamax} & \textbf{RMSprop} & \textbf{SGD} \\ \hline
\textbf{BRANCH} & 77.41\% & 78.69\% & 76.73\% & 53.95\% \\ \hline
\textbf{CLOCK} & 77.02\% & 80.74\% & 76.73\% & 52.37\% \\ \hline
\textbf{\begin{tabular}[c]{@{}c@{}}CYCLE\_ACTIVITY\end{tabular}} & 77.32\% & 80.42\% & 77.75\% & 54.59\% \\ \hline
\textbf{DATA} & 79.07\% & 80.01\% & 78.56\% & 53.86\% \\ \hline
\textbf{FLOPS\_DP} & 85.67\% & 86.90\% & 88.89\% & 58.75\% \\ \hline
\textbf{ICACHE} & 83.54\% & 89.02\% & 85.78\% & 55.44\% \\ \hline
\textbf{L2\_DATA} & 82.51\% & 86.00\% & 80.04\% & 53.14\% \\ \hline
\textbf{L2\_CACHE} & 80.31\% & 85.50\% & 82.27\% & 52.63\% \\ \hline
\textbf{L3\_DATA} & 82.85\% & 82.73\% & 84.30\% & 51.61\% \\ \hline
\textbf{L3\_CACHE} & 84.68\% & 85.40\% & 83.27\% & 53.61\% \\ \hline
\textbf{TLB\_DATA} & 96.34\% & 95.32\% & 95.36\% & 52.80\% \\ \hline
\textbf{TLB\_INSTR} & 48.90\% & 48.90\% & 48.90\% & 48.90\% \\ \hline
\textbf{UOPS} & 75.28\% & 78.86\% & 73.53\% & 52.12\% \\ \hline
\textbf{UOPS\_EXEC} & 76.90\% & 78.00\% & 76.77\% & 55.01\% \\ \hline
\textbf{UOPS\_ISSUE} & 77.46\% & 78.05\% & 78.17\% & 54.03\% \\ \hline
\textbf{UOPS\_RETIRE} & 78.31\% & 77.54\% & 77.96\% & 53.39\% \\ \hline
\end{tabular}
\end{table}

\begin{table}[!t]
\small
\centering
\caption{Accuracy with 80\% Training Dataset}
\label{tab:80}
\vspace{-0.1in}
\begin{tabular}{|c|c|c|c|c|}
\hline
\textbf{\begin{tabular}[c]{@{}c@{}}\end{tabular}} & \textbf{Adadelta} & \textbf{Adamax} & \textbf{RMSprop} & \textbf{SGD} \\ \hline
\textbf{BRANCH} & 74.88\% & 79.37\% & 77.43\% & 50.50\% \\ \hline
\textbf{CLOCK} & 79.50\% & 86.06\% & 79.37\% & 51.87\% \\ \hline
\textbf{\begin{tabular}[c]{@{}c@{}}CYCLE\_ACTIVITY\end{tabular}} & 77.75\% & 82.81\% & 78.11\% & 53.69\% \\ \hline
\textbf{DATA} & 77.12\% & 80.24\% & 77.61\% & 53.12\% \\ \hline
\textbf{FLOPS\_DP} & 87.00\% & 87.07\% & 90.51\% & 58.94\% \\ \hline
\textbf{ICACHE} & 83.12\% & 87.38\% & 87.38\% & 51.06\% \\ \hline
\textbf{L2\_DATA} & 81.43\% & 87.63\% & 81.88\% & 51.63\% \\ \hline
\textbf{L2\_CACHE} & 82.49\% & 86.50\% & 81.12\% & 52.68\% \\ \hline
\textbf{L3\_DATA} & 82.05\% & 87.00\% & 85.82\% & 52.81\% \\ \hline
\textbf{L3\_CACHE} & 84.87\% & 87.38\% & 85.06\% & 53.44\% \\ \hline
\textbf{TLB\_DATA} & 96.57\% & 95.52\% & 96.45\% & 52.37\% \\ \hline
\textbf{TLB\_INSTR} & 50.00\% & 50.00\% & 50.00\% & 50.00\% \\ \hline
\textbf{UOPS} & 72.88\% & 77.68\% & 73.75\% & 51.06\% \\ \hline
\textbf{UOPS\_EXEC} & 78.75\% & 77.68\% & 76.19\% & 52.12\% \\ \hline
\textbf{UOPS\_ISSUE} & 76.75\% & 79.93\% & 77.49\% & 53.31\% \\ \hline
\textbf{UOPS\_RETIRE} & 77.06\% & 78.81\% & 76.93\% & 51.94\% \\ \hline
\end{tabular}
\end{table}

\begin{table}[!ht]
\small
\centering
\caption{Accuracy with 90\% Training Dataset}
\label{tab:90}
\vspace{-0.1in}
\begin{tabular}{|c|c|c|c|c|}
\hline
\textbf{\begin{tabular}[c]{@{}c@{}}\end{tabular}} & \textbf{Adadelta} & \textbf{Adamax} & \textbf{RMSprop} & \textbf{SGD} \\ \hline
\textbf{BRANCH} & 76.00\% & 76.51\% & 78.12\% & 51.50\% \\ \hline
\textbf{CLOCK} & 77.49\% & 82.50\% & 77.49\% & 51.12\% \\ \hline
\textbf{\begin{tabular}[c]{@{}c@{}}CYCLE\_ACTIVITY\end{tabular}} & 79.24\% & 78.62\% & 79.76\% & 53.37\% \\ \hline
\textbf{DATA} & 76.38\% & 80.50\% & 76.25\% & 53.50\% \\ \hline
\textbf{FLOPS\_DP} & 85.50\% & 85.99\% & 90.88\% & 61.00\% \\ \hline
\textbf{ICACHE} & 84.75\% & 85.62\% & 84.13\% & 50.75\% \\ \hline
\textbf{L2\_DATA} & 81.38\% & 87.88\% & 83.75\% & 50.62\% \\ \hline
\textbf{L2\_CACHE} & 80.12\% & 86.50\% & 82.00\% & 53.62\% \\ \hline
\textbf{L3\_DATA} & 85.00\% & 83.50\% & 85.88\% & 50.88\% \\ \hline
\textbf{L3\_CACHE} & 87.14\% & 87.87\% & 85.88\% & 52.38\% \\ \hline
\textbf{TLB\_DATA} & 97.26\% & 96.52\% & 98.26\% & 51.00\% \\ \hline
\textbf{TLB\_INSTR} & 50.00\% & 50.00\% & 50.00\% & 50.00\% \\ \hline
\textbf{UOPS} & 72.14\% & 76.25\% & 75.37\% & 51.37\% \\ \hline
\textbf{UOPS\_EXEC} & 77.37\% & 78.74\% & 78.12\% & 51.88\% \\ \hline
\textbf{UOPS\_ISSUE} & 78.37\% & 78.86\% & 75.87\% & 53.25\% \\ \hline
\textbf{UOPS\_RETIRE} & 78.87\% & 76.50\% & 76.87\% & 51.37\% \\ \hline
\end{tabular}
\end{table}

\subsection{Performance Analysis of the ML Classifier}

As discussed in Section \ref{sec:methodology}.2 and Section \ref{sec:result}.3, we develop the predictive ML model by training with timeseries dataset of individual performance groups and associated metrics (see Table \ref{tab:eventgroup}) for given ransomware and goodware database. Note that selecting such groups and features depend on multiple factors: (1) inherent properties of the ML technique that utilizes such features to perform binary classification and (2) the program behavior that is running on the system (performing extensive encryption versus simple output printing). 

At first, we used {\fontfamily{pcr}\selectfont Keras} python library \cite{chollet2015keras} to implement the neural network under training. Equal distribution of benign and ransomware was maintained in the training dataset to prevent inclination of ML towards a specific dataset. The training was done on four different optimizers belonging to different classes, i.e., {\fontfamily{pcr}\selectfont SGD, Adamax, Adadelta,} and {\fontfamily{pcr}\selectfont RMSprop}, to calibrate network weights based on error for reducing the validation loss \cite{kerasoptimizer}. We used \si{25\%} of the training dataset for validation after each epoch to efficiently calibrate the loss function of the model. Also, to reduce any bias in the model due to misfitting, the accuracy analysis was performed over 50 iterations, where each run contained randomly shuffled executables and trained for 1000 epochs.

For an in-depth analysis of the RanStop technique, we analyzed the accuracy of the predictive model where it was developed using different sizes of training dataset, namely 70\% (Table \ref{tab:70}), 80\% (Table \ref{tab:80}), and 90\% (Table \ref{tab:90}) with previously mentioned optimizers. Here, each value represents the fraction of the total dataset that was used for training. The remainders of the dataset, i.e., 30\%, 20\%, and 10\% for respective cases, were used for testing. For each table, the detection accuracy (averaged over 50 iterations) is listed with the HPC groups (row) and optimizers (column). As seen from Tables \ref{tab:70}, \ref{tab:80}, and \ref{tab:90}, the {\fontfamily{pcr}\selectfont Adadelta} performed the best for all HPC groups in minimizing the over fitting. On the other hand {\fontfamily{pcr}\selectfont SGD} performed the worst. A detailed scrutiny suggested that the SGD overfitted the model due to the lack of data endpoints. The rest three of the optimizers performed quite similar to each other. On the other hand, the most prominent micro-architectural event group was {\fontfamily{pcr}\selectfont TLB\_DATA} which provided the best detection rate, whereas {\fontfamily{pcr}\selectfont TLB\_INSTR} provided the worst outcome. The results show that even with 70\% training dataset, Ranstop was able to achieve as high as \si{96\%} accuracy by correctly identifying benign versus crypto-ransomware. And the accuracy goes to as high as \si{97\%} while the training dataset contains \si{90\%} of the total timeseries data. We also note that the programs (ransomware/benign) used for testing the model were not any part of the training dataset, as mentioned previously. Therefore, this supervised classifier is fully compatible for detecting unknown ransomware, i.e., emerging variants with no (or, very limited, if needed at all) retraining.

False Negative Rate and False Positive Rate are also calculated using the equation \ref{eq:FN}. We consider both false negatives and false positives as major drawbacks for any ransomware (or malware, in general) detection technique, since false positives, i.e., true benign programs deemed as ransomware, cause inconvenience and probable denial of service, whereas false negatives, i.e. true ransomware detected as benign program, can cause catastrophic damage to the system. 

\vspace{-0.05in}
\begin{equation}\label{eq:FN}
FN Rate = \frac{FN}{FN+TP}, ~ FP Rate = \frac{FP}{FP+TN}
\end{equation}
\vspace{0.05in}

Tables \ref{tab:adadelta} and \ref{tab:rmsprop} show the results for statistical metrics (True Positive, True Negative, False Positive, and False Negative) for two of the best performing optimizers, i.e. Adadelta and RMSprop. The results are again averaged over 50 iterations for each performance group to remove any bias or residual count. The results show that if we remove the outlier case of {\fontfamily{pcr}\selectfont TLB\_INSTR}, the false negative (identifying crypto-ransomware as benign) rate is less then \si{1\%} for many of the performance groups. The result also shows the false positive (identifying benign as ransomware) is little high that can be concluded to the fact that many micro-architectural activities of benign programs may resemble that of a crypto-ransomware. We expect that the false positive will significantly reduce with the increase in the dataset size and diversity as a future work. 

\begin{table}
\small
\centering
\caption{Statistics for Adadelta with 70\% training data}
\label{tab:adadelta}
\vspace{-0.1in}
\begin{tabular}{|c|c|c|c|c|}
\hline
\textbf{} & \textbf{TP} & \textbf{TN} & \textbf{FP} & \textbf{FN} \\ \hline
\textbf{BRANCH} & 49.11\% & 29.57\% & 19.36\% & 1.96\% \\ \hline
\textbf{CLOCK} & 48.09\% & 32.64\% & 16.30\% & 2.98\% \\ \hline
\textbf{CYCLE\_ACTIVITY} & 50.51\% & 31.62\% & 19.45\% & 0.55\% \\ \hline
\textbf{DATA} & 48.09\% & 31.91\% & 17.02\% & 2.98\% \\ \hline
\textbf{FLOPS\_DP} & 47.49\% & 39.40\% & 9.53\% & 3.57\% \\ \hline
\textbf{ICACHE} & 48.21\% & 40.81\% & 8.13\% & 2.85\% \\ \hline
\textbf{L2\_DATA} & 47.87\% & 38.13\% & 10.81\% & 3.19\% \\ \hline
\textbf{L2\_CACHE} & 50.09\% & 35.40\% & 13.53\% & 0.98\% \\ \hline
\textbf{L3\_DATA} & 48.00\% & 34.72\% & 14.21\% & 3.06\% \\ \hline
\textbf{L3\_CACHE} & 46.00\% & 39.40\% & 9.53\% & 5.06\% \\ \hline
\textbf{TLB\_DATA} & 49.57\% & 45.74\% & 3.19\% & 1.49\% \\ \hline
\textbf{TLB\_INSTR} & 0.00\% & 48.94\% & 0.00\% & 51.06\% \\ \hline
\textbf{UOPS} & 49.02\% & 29.83\% & 19.11\% & 2.04\% \\ \hline
\textbf{UOPS\_EXEC} & 50.89\% & 27.11\% & 21.83\% & 0.17\% \\ \hline
\textbf{UOPS\_ISSUE} & 49.83\% & 28.21\% & 20.72\% & 1.23\% \\ \hline
\textbf{UOPS\_RETIRE} & 50.30\% & 27.23\% & 21.70\% & 0.77\% \\ \hline
\end{tabular}
\end{table}

\begin{table}[t]
\small
\centering
\caption{Statistics for RMSprop with 70\% training data}
\label{tab:rmsprop}
\vspace{-0.1in}
\begin{tabular}{|c|c|c|c|c|}
\hline
\textbf{} & \textbf{TP} & \textbf{TN} & \textbf{FP} & \textbf{FN} \\ \hline
\textbf{BRANCH} & 48.77\% & 27.96\% & 20.98\% & 2.30\% \\ \hline
\textbf{CLOCK} & 49.15\% & 27.57\% & 21.36\% & 1.91\% \\ \hline
\textbf{CYCLE\_ACTIVITY} & 50.04\% & 29.36\% & 21.70\% & 1.02\% \\ \hline
\textbf{DATA} & 49.62\% & 28.94\% & 20.00\% & 1.45\% \\ \hline
\textbf{FLOPS\_DP} & 48.64\% & 40.26\% & 8.68\% & 2.43\% \\ \hline
\textbf{ICACHE} & 47.87\% & 37.91\% & 11.02\% & 3.19\% \\ \hline
\textbf{L2\_DATA} & 44.81\% & 35.23\% & 13.70\% & 6.26\% \\ \hline
\textbf{L2\_CACHE} & 48.26\% & 34.00\% & 14.94\% & 2.81\% \\ \hline
\textbf{L3\_DATA} & 46.77\% & 37.53\% & 11.40\% & 4.30\% \\ \hline
\textbf{L3\_CACHE} & 46.77\% & 36.51\% & 12.43\% & 4.30\% \\ \hline
\textbf{TLB\_DATA} & 50.17\% & 45.19\% & 3.74\% & 0.89\% \\ \hline
\textbf{TLB\_INSTR} & 0.00\% & 48.94\% & 0.00\% & 51.06\% \\ \hline
\textbf{UOPS} & 48.38\% & 25.15\% & 23.79\% & 2.68\% \\ \hline
\textbf{UOPS\_EXEC} & 49.96\% & 26.81\% & 22.13\% & 1.11\% \\ \hline
\textbf{UOPS\_ISSUE} & 50.43\% & 27.74\% & 21.19\% & 0.64\% \\ \hline
\textbf{UOPS\_RETIRE} & 50.04\% & 27.91\% & 21.02\% & 1.02\% \\ \hline
\end{tabular}
\end{table}

In Table \ref{tab:compare}, we provide a comparative analysis between our proposed RanStop scheme and existing state-of-the-art techniques for ransomware detection. The table lists different detection techniques, dataset sizes, and performance metric. As it is shown, RanStop has provided signficiantly better result over a comprehensive database of ransomware and goodware; and can provide an early detection with very high accuracy.

\begin{table*}[t]
\small
\centering
\caption{Comparative Analysis of Existing Techniques}
\label{tab:compare}

\begin{tabular}{|l|l|l|l|l|l|}
\hline
                             & \begin{tabular}[c]{@{}l@{}}Kharazz\\ et al. {[}11{]}\end{tabular} & \begin{tabular}[c]{@{}l@{}}Scaife\\ et al. {[}6{]}\end{tabular} & \begin{tabular}[c]{@{}l@{}}Moussaileb\\ et al. {[}17{]}\end{tabular} & \begin{tabular}[c]{@{}l@{}}Alam\\ et al. {[}16{]}\end{tabular} & \textbf{RanStop}                   \\ \hline
Detection Type               & Static                                                            & Static                                                          & Static + Dynamic                                                     & Dynamic                                                        & Dynamic                            \\ \hline
Key Features                 & -                                                                 & API Calls                                                       & graph                                                                & hardware-assisted                                              & hardware-assisted                  \\ \hline
Ransomware Database Size     & -                                                                 & -                                                               & $\sim$700                                                            & \textless{}5                                                   & \textbf{80}                        \\ \hline
Goodware Program Versatility & Random                                                            & Random                                                          & Random                                                               & Random + Computationally Intensive                             & Random + Computationally Intensive \\ \hline
Signature Collection Time    & -                                                                 & -                                                               & -                                                                    & In order of seconds                                            & \textbf{In order of milliseconds}  \\ \hline
Average Accuracy             & -                                                                   & -                                                                 & in order of 60\%                                                     & 100\% (for selective RW)                                       & 97\%                               \\ \hline
\end{tabular}
\end{table*}

\section{Conclusion} \label{sec:conclusion}
In this paper, we present a hardware-assisted crypto-ransomware runtime detector, called RanStop. Our proposed technique can detect ransomware with an average of \si{97\percent} with data collected as early as \si{2\milli\second} from the start of execution of a ransomware. Such a very early detection technique ensures that a system, even if somewhat infected, will suffer little or no damage by crypto-ransomware, and, therefore, robust against any stealthy ransomware attack. The LSTM-based ML modeling scheme offers a high accuracy for multiple optimizers; giving the user complete freedom to choose for while deploying the model for runtime detection. Although fast and accurate, our proposed scheme, like many other existing techniques, suffers from additional challenges -- especially, (1) Like any other watchdog program, the runtime detection program itself may be vulnerable to malicious infections and the obtained hardware data may get corrupted. Hence, the program must run at the highest privilege level, be independent of any other program, and have access to sufficient resources (e.g., physical memory). (2) Due to hardware (physical) limitations, it is not possible to simultaneously access and monitor all micro-architectural events via available limited number of HPCs. Moreover, the micro-architectural events collected by HPCs are historically performance-oriented. Therefore; it does not necessarily provide security-aware micro-architectural events that may become significant for the detection of emerging threats. This will require development and implementation of additional HPC-like register for wider event coverage. We leave this challenges as a future scope to this work. 

\bibliographystyle{IEEEtran} 
\bibliography{Bib_ACSAC2019_Ransomware}

\end{document}